\documentclass[a4paper,12pt]{article}

\usepackage{amsmath,amssymb,amsthm}
\usepackage{latexsym,graphicx,color,bbm,subfigure}
\usepackage{pdfsync}   
\usepackage{bbm}
\usepackage{color}

\newcommand{\be}{\begin{equation}}
\newcommand{\ee}{\end{equation}} 
\newcommand{\beq}{\begin{eqnarray}}
\newcommand{\eeq}{\end{eqnarray}}

\newcommand{\p}{\partial}
\newcommand{\Tr}{{\rm Tr}}

\newcommand{\bea}{\begin{eqnarray}}
\newcommand{\eea}{\end{eqnarray}}
\def\Tr{ \hbox{\rm Tr}}

\def\tr{ \hbox{\rm tr}}
\def\de{\partial}

\def\bra{\langle}
\def\ket{\rangle}

\begin{document}

\thispagestyle{empty}

\begin{flushright}
IFUP-TH/2014-16
\end{flushright}
\vspace{10mm}

\begin{center}
{\huge \bf 
Monopole-vortex complex at large distances and   nonAbelian duality
 } 
\\[15mm]
{}
{Chandrasekhar Chatterjee}$^{a,b}$ \footnote{\it e-mail address:
chatterjee.chandrasekhar(at)pi.infn.it},  
{Kenichi Konishi}$^{b,a}$ \footnote{\it e-mail address:
konishi(at)df.unipi.it}, 

\vskip 6 mm

\bigskip\bigskip
{\it

  INFN, Sezione di Pisa $^{a}$,
Largo Pontecorvo, 3, Ed. C, 56127 Pisa, Italy 
\\
~Department of Physics ``E. Fermi'', University of Pisa  $^{b}$, \\
Largo Pontecorvo, 3, Ed. C, 56127 Pisa, Italy

 }

\end{center}

\vskip 6 mm

\bigskip
\bigskip

\begin{center}

{\bf Abstract}\\[5mm]
{\parbox{14cm}{\hspace{5mm}
\small

We discuss the large-distance approximation of the monopole-vortex complex soliton in a hierarchically broken  gauge system, $SU(N+1) \to SU(N)\times U(1) \to {\bf 1}$, 
in a color-flavor locked $SU(N)$ symmetric vacuum. The ('t Hooft-Polyakov) monopole of the higher-mass-scale breaking appears as a point and acts as a source of the thin vortex generated by the lower-energy gauge symmetry breaking.  The exact color-flavor diagonal symmetry of the bulk system  is broken by each individual  soliton,  leading  to  nonAbelian orientational $CP^{N-1}$  zeromodes propagating in the vortex worldsheet, well studied in the literature.  But  since the vortex ends at the monopoles these fluctuating modes    
endow the monopoles with a  local $SU(N)$ charge.  This phenomenon is  studied by performing the duality transformation in the presence of the $CP^{N-1}$  moduli space.
The effective action is a $CP^{N-1}$ model defined on a finite-width worldstrip.  

}}
\end{center}

\newpage
\pagenumbering{arabic}
\setcounter{page}{1}
\setcounter{footnote}{0}
\renewcommand{\thefootnote}{\arabic{footnote}}

\section{Introduction}

The idea of  nonAbelian monopoles and the associated concept of nonAbelian duality  has proven to be peculiarly elusive. The exact Seiberg-Witten solutions and its generalizations  in the context of ${\cal N}=2$ supersymmetric theories \cite{SW1}-\cite{SUN}  show that  the massless monopoles appearing at various simple singularities of  the quantum modiuli space of vacua
(QMS) are Abelian. There are clear evidences \cite{SW1}-\cite{KT} that they are indeed the 't Hooft-Polyakov monopoles becoming light by quantum effects.  This has (mis-)led some people  to believe that the monopoles seen in the low-energy dual theories of the soluble ${\cal N}=2$ models are always Abelian. 
Actually, in many degenerate singularities  where Abelian vacua coalesce   (e.g., at certain values of the bare masses and/or at special points of the vacuum moduli space),   nonAbelian monopoles regularly make appearance \cite{APS}-\cite{CKM} as dual, massless degrees of freedom.  They  correctly describe the infrared phenomena such as confinement and  global symmetry breaking \footnote{A typical example is the so-called $r$-vacua \cite{APS}-\cite{CKM}  of ${\cal N}=2$,  $SU(N)$ SQCD with $N_f$ quarks in the equal mass limit, where an exact flavor $SU(N_f)$ symmetry and the dual $SU(r)$ ($r < N_f/2$) gauge symmetry appear simultaneously.  Evidently the correct realization of the global symmetry ($SU(N_{f}$)) and renormalization group flow ($r < N_f/2$)  are interrelated subtly.}. NonAbelian monopoles are also known to appear in infrared-fixed-points of ${\cal N}=1$
supersymmetric theories, those in Seiberg's conformal window (${\cal N}=1$  SQCD,   for  $\tfrac{3}{2} N_{c}< N_{f} < 3 N_{c}$) \cite{SeibergDual} being the most celebrated examples 
\footnote{There is a good reason for the frequent  appearance of nonAbelian monopoles in the nontrivial fixed-point conformal theories.   Due to renormalization-group flow, non-Abelian monopoles tend to interact too strongly in the infrared:  
unless they do not acquire sufficiently large flavor content, they would not be seen in the infrared. Abelian monopoles are infrared free, so they can appear more easily as infrared degrees of freedom.
 The limit  $r\le  N_f/2$  for the $r$-vacua is a manifestation of this fact.  The nontrivial IFPT  conformal theories are the critical cases: nonAbelian monopoles and quarks  appear together as low-energy massless degrees of freedom.  
  }.  Even more intriguing is the situation in highly singular vacua occurring in the context of ${\cal N}=2$  SQCD  for some critical quark mass in $SU(N)$ theories \cite{SCFT,SimoneLorenzo}   or in the  massless limit of  $SO(N)$ or $USp(2N)$ theories \cite{APS,Eguchi,CKKM}.  
These infrared-fixed-point SCFT's   are described by a set of relatively nonlocal and strongly-coupled monopoles and dyons, and it is quite a nontrivial matter  to analyze what happens when a ${\cal N}=1$ deformation is introduced in the theory.  Is the system brought into confinement phase? And if so, how is it described?  Only quite recently some  progress was made concerning these questions \cite{SGKK},   taking full advantage of  some beautiful results of Argyres, Gaiotto, Seiberg, Tachikawa and others \cite{Argyres:2007cn,GST,Simone}.  
Confinement vacua near such highly singular  SCFT exhibit interesting features which could provide important hints about the not-yet-known confinement mechanism in QCD. 

Leaving aside this deep issue here, the point is that appearance of  the nonAbelian monopoles  as infrared degrees of freedom is quite common in strongly interacting nonAbelian gauge theories. 
What is lacking still is the understanding of these quantum objects, or in other words,  of their semi-classical origin. 
This question is a relevant one, in view of the difficulties associated with the straightforward idea of semi-classical nonAbelian monopoles \cite{CDyons,EW,DFHK}.

 It is our aim in this paper to take a few more  steps towards  
 elucidating the mysteries of nonAbelian duality.    For this purpose we study a system with hierarchical symmetry breaking \footnote{Although for definiteness we here consider $SU(N)$ theories only, the idea of hierarchical symmetry breaking and the monopole-vortex connection in a color-flavor locked vacuum can naturally be extended to other gauge theories such as $SO(2N)$ or $USp(2N)$
  \cite{GK, GJK}.  Such an extension is straightforward but interesting: the monopoles transform according to spinor representations of the dual $Spin(2N)$ or   $SO(2N+1)$, respectively, in these cases}
 \be
SU(N+1)_{\rm color} \otimes SU(N)_{\rm flavor} 
  \stackrel{v_{1}}{\longrightarrow} 
(SU(N)\times U(1))_{\rm color}  \otimes SU(N)_{\rm flavor} 
\stackrel{v_{2}}{\longrightarrow} SU(N)_{C+F} \ ,   \label{SUNhierarchy}
\ee
with
\be  
v_{1} \gg v_{2}\ ,  
\ee
as in \cite{ABEKY}-\cite{Konishi}.  The homotopy group associated with the gauge symmetry breaking, 
\be \Pi_2(SU(N+1)/SU(N)\times U(1)) \sim {\mathbbm Z}     \label{homot1}\ee
supports  monopoles with quantized magnetic charges, whereas 
the low-energy $U(N)$ symmetry breaking  with 
\be \Pi_1(SU(N)\times U(1))  \sim {\mathbbm Z}     \label{homot2}     \ee
 implies vortices.  As neither of them exists in the full theory,   \be \Pi_2(SU(N+1)) = \Pi_1(SU(N+1))= {\mathbbm 1},\ee
the vortex must  end: the endpoints are the monopoles.  This fact can be rephrased by the short exact sequence of the associated homotopy groups:
\be  {\mathbbm 1} =  \Pi_{2}(SU(N+1)) \to     \Pi_{2}(\tfrac{SU(N+1)}{SU(N)\times U(1)})   \to  \Pi_{1}(SU(N)\times U(1))  \to   \Pi_{1}(SU(N+1))    =  {\mathbbm 1}\;.\label{connection}
\ee
The fact that neither monopole or vortex exist as  stable solitons of the full theory does not prevent us from investigating these configurations. The  idea is to keep the mass-scale hierarchy $v_{1} \gg v_{2}$ as strong as we wish, so that the concept of monopole or vortex is as good as any approximation used in physics \footnote{Of course, quantization of the radial and rotational motions of the monopoles can stabilize such a system dynamically, without need of the hierarchy,
$v_{1} \gg v_{2}$.}.  

Note that the topological classifications such as  Eqs.~(\ref{homot1}), (\ref{homot2}), are not fine enough to specify  the minimum-energy configuration in each class. There are continuously infinite degeneracy of such minimum configurations due to the breaking of the exact global color-favor $SU(N)_{C+F}$ symmetry by  the vortex and  monopole.  The connection implied by Eq. (\ref{connection})  however means that each minimal vortex configuration  of the minimum class of  Eq.  (\ref{homot2})  ends at a monopole of the miminum class  Eq. (\ref{homot2}).  This connection endows the monopole with the same  
$CP^{N-1}$ orientation moduli of the nonAbelian vortex. 

The new $SU(N)$ quantum number of the monopole arises as the isometry group of this $CP^{N-1}$ moduli space, following from 
the exact color-flavor symmetry of the original gauge system.  The monopole transforms in the fundamental representation of this $SU(N)$. Fluctuation of the monopole $SU(N)$ charge  excites the well-known non-Abelian vortex zero modes \cite{Hanany:2003hp,ABEKY}, which propagate as massless particles in the 2D worldsheet. One way of thinking about this result is that {\it  the non-normalizable  $3D$  gauge zero modes of the monopole, when dressed  by flavor charges,  turn  into normalizable $2D$ modes on the vortex world sheet}.

The new $SU(N)$ symmetry is a result of the color-flavor combined transformations acting on the soliton monopole-vortex configurations:  the latter  is a nonlocal field transformation. 
Nevertheless, in the dual description the new $SU(N)$ acts locally on the monopole. 
This is typical of electromagnetic duality. 

This $SU(N)$ charge of the monopole is a confined charge, as the excitation does not propagate outside the monopole-vortex-antimonopole complex. The $M-V-{\bar M}$ complex is a singlet as a whole. 
The monopoles appear as confined objects, the vortex playing the role of the confining string. This is correct as the original $SU(N)$ gauge system is in a completely broken, Higgs phase.  The dual system must be in confinement phase.

  The following is an attempt to make these ideas a little more concrete.

\section {The model}

Our aim is to study a simplest possible model which realizes the hierarchical  symmetry breaking  Eq.(\ref{SUNhierarchy})  in which 
 the light  matter and gauge fields  interact  with the monopole arising from the higher-mass symmetry breaking.   The action can be taken in  the form, 
\be 
\mathcal{L} = 
-\frac{1}{4} \left(F_{\mu\nu}\right)^2 
+    |{\cal D}_{\mu} \phi|^2 
+\sum_{I=1}^{N_{F}}\left|{\cal D}_{\mu} q_{I}\right|^2 
- V(\phi, q) \, ,\label{Themodel}  \ee
where $\phi$ is a scalar field in the adjoint representation of $SU(N+1)$, $q_{I}$, where $I=1,2,\ldots, N_{F}=N$,    are a set of other scalar fields, in the fundamental representation. Inspired by the ${\cal N}=2$ 
theories,  we take   
\be
V   = 
\sum_A \left|\mu \, \phi^A + \lambda  \, {q}_I^{\dagger} \, T^A q^I \right|^2  
+ \sum_{I, i}\left |  ( T^A  \phi^A+ m_I   )_{ij}   \, q_j^I  \right|^2 \,.   \label{potential} 
\ee
where $m_1, m_2, \ldots m_{N}$  are the (bare)  masses of  the scalar fields $q$, and $\mu \ll |m_{I}|$.   The quartic coupling $\lambda$ does not play a particular role in our discussion below, and will be set to unity. 
In order to attain the minimum of  the potential, 
$V=0$,   the scalar field $q_I$ is either a non vanishing eigenvector of the $\phi$ with eigenvalue, $m_I$, or   must vanish.  We shall take the equal mass limit, $m_1= m_2= \ldots = m_{N}=m_0$ and choose to work in  the vacuum with
\be      \bra \phi \ket    =  \bra \phi^{A} T^{A} \ket  =m_{0}\,     \left(\begin{array}{cc}   N   &  \\ &  -{\mathbbm 1}_N   \end{array}    \right)  \label{phivev}  \ee
breaking the  $SU(N+1)$ gauge symmetry to  $SU(N) \times U(1)$.   An inspection of the second term of the potential shows that  the first (color)  component of the scalars $q_{I}$  becomes massive
for all $I$ (with vanishing VEV), with mass  \be v_{1} \equiv  m_{0} (N+1),  \ee
and decouples
at mass scales lower than  that.  
The other components  are nontrivial eigenvectors  $q_I$ of   $\phi$.   $N_f=N$  eigenvectors can be taken to be orthogonal to each other,  $ \bra  q_{I}^{a} \ket  =  c_I \,  \delta^{a}_{I}.$  The first term tells
\be   \tr  \,   t^{A}   \sum_{I}  (\, q_{I}  \,  q^{\dagger\, I} ) = 0\;,  \qquad  t^{A} \subset SU(N),  
\ee
that is,   $ \sum_{I}  q_{I}  \,  q^{\dagger\, I}  \propto {\mathbbm 1}_{N}$. In other words, all $c_I$'s  are equal. Their normalization is fixed by the $A=0$  (see Eq.~(\ref{fixed})) term to be 
\be      \bra  q_{I}^{a} \ket  =   v_{2} \,  \delta^{a}_{I}\;, \qquad          v_{2} \equiv   \sqrt{ 2 (N+1)  \mu m_{0}}  \ll v_{1}\,. 
\ee
showing that the gauge symmetry is completely broken at low energies, leaving however the  color-flavor diagonal symmetry $SU(N)_{C+F}$ unbroken.

 \section{Point:   the monopole \label{macro}} 

Let us write the VEV of $\phi$ as  
\be  \phi(x)= v_{1}\, M(x), \qquad    \bra M \ket    =    \sqrt {\frac{2N}{N+1}} \,  T^{(0)}, \qquad   T^{(0)}=  \frac{1}{\sqrt{2N(N+1)}}  \left(\begin{array}{cc}   N   &  \\ &  -{\mathbbm 1}_N   \end{array}\right) \;.    \label{fixed}\ee
  As the system admits the topological defect (the monopole),  we need to retain the degrees of freedom corresponding to the nontrivial winding
\[  \Pi_{2}(SU(N+1)/U(N))  = {\mathbbm Z}   
\]
that is,
  \be         M(x)   =  \sqrt {\frac{2N}{N+1}} \,   U(x)  \, T^{(0)}\,    U^{\dagger}(x), \qquad   \Tr \,  (T^{(0)})^{2} =\frac{1}{2}\;. \label{UtoM}
\ee
The $M(x)$ ($U(x)$) field  defines the direction of the symmetry breaking, 
\be   SU(N+1)/U(N) \sim CP^{N},   \label{symbreak}
\ee 
and can be expressed by a complex $(N+1)$-component vector field $z(x)$ as 
\be  M =   z\, {\bar z}  - \frac{1}{N+1} {\mathbbm 1},   \qquad  {\bar z}\,  z =1.    \label{Mfiel}  \ee
By introducing the $N+1$ orthonormal eigenvectors of $M$, $z(x)$ and $e_{i}(x)$ ($i=1,2,\ldots N$)   with eigenvalues, \[\frac{N}{N+1},  \quad- \frac{1}{N+1}, \quad- \frac{1}{N+1},\quad\ldots, \quad - \frac{1}{N+1},\]
$U(x)$ can be written as 
 \be U(x)  =  \left(    \left(\begin{array}{c}   \\ z  \\  \\  \end{array}\right)   \left(\begin{array}{c}   \\ e_{1}  \\  \\  \end{array}\right)   \cdots   \left(\begin{array}{c}   \\ e_{N}  \\  \\   \end{array}\right)   \right).    \label{winding}
\ee
The $z^{a}$ and $e^{a}_{i}$ can be thought of as local vielbeins \cite{EGKM}.

The gauge field can be taken so as to satisfy the so-called Cho gauge \cite{Cho}
\be  ( D_{\mu}\phi )/ v_{1}=  D_{\mu}M =   \de_{\mu} M  - i  g  \, [ A_{\mu}, M]  =   0\;,  \label{ChoCond}
\ee
which amounts to the low-energy (large distance) approximation where the monopole appears as a point. Namely,  
only the winding directions, $SU(N+1)/SU(N)\times U(1)\sim CP^N$,   are  kept as the dynamical degrees of freedom associated with the monopole.    
Explicitly, we take the  gauge field in  the form, 
\be   A_{\mu} =   C_{\mu} \, M(x) +  \frac{i}{g} \, [M(x), \de_{\mu}M(x) ]  +  B_{\mu},     \label{Ansatz}  
\ee
where
\be   { B}_{\mu\, b}^{a} =  \sum_{i,j =1}^N e_{i}^{a}  b^{i}_{\mu\, j}  {\bar e}^{j}_{b}\;,  \qquad a,b=1,2,\ldots, N+1\;.
\ee
$ C_{\mu} $  is the Abelian gauge field in the direction of the scalar VEV,   $b^{i}_{\mu\, j} $  are the components of the gauge fields of the "unbroken"  $SU(N)$, and $i\, [M(x), \de_{\mu}M(x) ]/g$ represents the monopole field. 
It can be easily checked that the connection Eq. (\ref{Ansatz}) indeed satisfies the gauge condition Eq.~(\ref{ChoCond}).

Note that the Cho condition Eq. (\ref{ChoCond}) does not  uniquely determine  the component of the gauge field $A_{\mu}$  orthogonal to $z z^{\dagger}$,    as\[     [ e (\ldots)  {\bar e}, M] =0\;. 
\]
In Eq.~(\ref{Ansatz}) we have chosen $A_{\mu}$ so that it contains precisely the monopole configuration \footnote{The connection Eq. (\ref{Ansatz}) in fact  differs from the one discussed in \cite{EGKM}  by a term of the form,  $e (\ldots ) {\bar e}$,  more precisely by  
\[   E_{\mu\, b }^{a}   =  i \, e_{i}^{a} \,  \left[ {\bar e}^{i}_{c} \de_{\mu} e_{j}^{c} -  \delta^{i}_{j}\frac{1}{N}  \Tr  ({\bar e} \, \de_{\mu} e)\right]  \, {\bar e}^{j}_{b}\;.\]
This term was subtracted from $A_{\mu}$  in  \cite{EGKM}, in order to keep  the monopole term of the connection invariant under the ``unbroken''  $SU(N)$ gauge transformation.  
For the purpose of the present paper of studying {\it how the monopole transforms}  (in a fixed gauge)   it is not only unnecessary but somewhat misleading to make such a rearrangement of the gauge field. 
The monopole simply resides in a broken $su(2) \subset su(N+1)$, whose embedding direction 
is locked with the low-energy vortex direction in color and flavor, and they transform together.  See  below.  } 
\be    i\, [M(x), \de_{\mu}M(x) ]   \subset   su(2)  \subset   su(N+1),
\ee
besides the "unbroken" massless $SU(N)$ gauge fields  $b_{\mu \nu}$ and the Abelian gauge field $C_{\mu}$  (the $U(1)$ rotations around $M$ direction).  

The transformation properties of the gauge connection above have been studied in \cite{EGKM}.
Under the U($1$) transformation around the $M$ direction,
\be  U =  e^{i \alpha M}  =    e^{i \alpha  \frac{N}{N+1} }  z {\bar z}   +   e^{- i \alpha  \frac{1}{N+1} }    e {\bar e}\;
\label{u1tr}\ee
the Abelian field $C_{\mu}$ transforms as expected: 
\be  C_{\mu} \to  C_{\mu}-  \,\partial_{\mu}\alpha\;.  \label{effU1tr}\ee
Under the  $SU(N)$ transformations commuting with $M$,
\begin{equation}  U=  \exp(i \,\omega^{\textsf{A}} e \, t^{\textsf{A}}\bar{e})=
e\, \Omega \, \bar{e}+z\bar{z}\;, \qquad  \Omega =   \exp(i\, \omega^{\textsf{A}}t^{\textsf{A}})
\;.   \end{equation}
where  $(t^{\textsf{A}})^{i}_{j}$  ($i,j =1,2,\ldots N$) are  $SU(N)$  generators,  
$b_{\mu}$ transforms as the usual nonAbelian gauge field: 
\be b_{\mu}  \to    b^{U}_{\mu}=\Omega \, b_{\mu} \Omega^{\dagger}-i\,\partial_{\mu}\Omega \, \Omega^{\dagger}\;. \label{localBtr}  \ee

\subsection{The minimal monopoles}

For the minimal monopole   one can choose   $z$ and one of the $e$'s  to live in a $SU(2) \in SU(N+1)$ subgroup  
and consider their winding only \cite{EGKM},
\be  z=\left(\begin{array}{c}     \cos \frac{\theta}{2} \\ 0  \\\vdots  \\  e^{i \varphi}  \sin \frac{\theta}{2}   \\  0 \\    \vdots     \end{array}\right)\;,  \qquad 
 e_i =\left(\begin{array}{c}     -  e^{-i \varphi}   \sin \frac{\theta}{2} \\ 0  \\\vdots  \\   \cos  \frac{\theta}{2}   \\  0 \\    \vdots     \end{array}\right)   \;;  \label{minimonopole}
\ee
\be   e_j ^a=  \delta_{j+1}^{a}, \qquad  a=1, \ldots N+1, \quad   j=1,\ldots, N, \quad j \ne i\;.
\ee
  In other words, only  the vielbeins $z$ and $e_i $ in  the  first    and $ (i+1)$-th    color components  are relevant for the monopole.  
    They take the form of  normalized spin $1/2$ wave functions,  spin  up for $z$, spin down for $e_i$,  all the rest of the components being zero.    Other vielbeins $e_j$, $j \ne i$, have canonical, orthonormal unit vector forms. 
   $z$ and $e$'s are not independent but are related by the completeness and orthonormality conditions
\be     z^a  {\bar z}_b +   \sum_{i=1}^{N} e_i^a  {\bar e}^i_b =  \delta^a_b,   \qquad {\bar z}_a  z^a = 1, \quad   {\bar e}^i_a e_j^a=\delta^i_j\;, \quad  {\bar z}_a  e^a_i=0\;. 
\ee
but it is an arbitrary choice which of the vielbeins $e$ winds together with $z$. 

For the minimal monopole, Eq.  (\ref{minimonopole}),
the $(N+1)\times (N+1)$  matrix field $M$ takes the form, 
\be   M= \phi/v_1 =     z\, {\bar z}  - \frac{1}{N+1} {\mathbbm 1}  =    \frac {{\bf n} \cdot  \tau} {2}  + \frac {{\bf 1}_{1, i+1} }{2}  -   \frac{ {\bf 1}   }{N+1}, \ee
\be   {\bf n}= \frac{\bf r}{r} = (\sin \theta \cos \varphi, \sin \theta \sin \varphi, \cos \theta)\;,  \label{minimonopole1} \ee
where $\tau$ and $ {\bf 1}_{1, i+1} $    are the Pauli matrices and the  $2 \times 2$ unit matrix   in the $(1, i+1)$ subspace.   The monopole part of the gauge field Eq. (\ref{Ansatz})  is simply: 
\be A_{\mu}^{(monopole)} =  \frac{i}{g}  [ M, \de_{\mu} M  ]=   - \frac{\tau }{2 g }  \cdot  ( {\bf n} \times  \de_{\mu} {\bf n} ) \;, \label{minimonopole2}
\ee
which is indeed the singular Wu-Yang monopole lying in the  $su(2)\subset  su(N+1)$ algebra  in the $(1, i+1)$ subspace. This is nothing but the asymptotic form of the  't Hooft-Polyakov monopole, far from its center ($R \gg 1/v_{1}$).

Rotating this monopole field in one of the legs ($i=1,2,\ldots, N$), i.e.,  in the "unbroken" $SU(N)$  group,  amounts to the straightforward  idea of nonAbelian monopoles: a set of 
configurations of degenerate  mass,  which apparently  belong to the fundamental representation of  the $SU(N)$.   
A closer examination however reveals  the well-known difficulties (e.g.,  the topological obstruction \cite{CDyons}).
Any deeper understanding of the non-Abelian monopole notion  necessarily involves  an exact  flavor symmetry, as is fairly well known, and our following discussion is precisely based on such a consideration.

 The gauge field tensor can be calculated straightforwardly:
\bea  (F_{\mu \nu})^a_b  &=&      
\left(  \de_{\mu} A_{\nu} - \de_{\nu} A_{\mu}    - i g \, [A_{\mu}, A_{\nu}]   \right)^a_b  \\
&=&   z^a\, \{ \frac{M_{\mu \nu}}{2} +   \frac{N}{N+1} C_{\mu \nu} \}   {\bar z}_b  + e^a_i \, \{ (K_{\mu \nu})^i_j -  \frac{1}{N+1} C_{\mu \nu} \delta^i_j +   (b_{\mu \nu})^i_j  +  (h_{\mu \nu})^i_j  
 \}   \,  {\bar e}^j_b     \nonumber  
\eea 

where 
\be   C_{\mu \nu} =   \de_{\mu} C_{\nu} -\de_{\nu} C_{\mu},
\qquad    M_{\mu \nu} \equiv   \de_{\mu} N_{\nu} -\de_{\nu} N_{\mu},   \qquad  N_{\mu}\equiv   \frac{ 2  i}{g} \, {\bar z}_{a} \de_{\mu} z^{a}\;;   \label{gfield1}
 \ee 
\be   (K_{\mu \nu})^i_j   =  \de_{\mu}  (P_{\nu})^i_j -\de_{\nu}  (P_{\mu})^i_j  -i \,g \,  [P_{\mu}, P_{\nu}]^i_j\;,      \qquad  (P_{\mu})^i_j=    \frac{i}{g} \, {\bar e}^i_a \de_{\mu} e^a_j \;; \label{gfield2}
 \ee
\be   b_{\mu \nu} =   \de_{\mu} b_{\nu} -  \de_{\nu} b_{\mu} - i \,g \, [ b_{\mu}, b_{\nu} ]\;;   \qquad 
 h_{\mu \nu} = -i g  \left( [P_{\mu }, b_{\nu} ] -  [P_{\nu }, b_{\mu} ]  \right)\;.   \label{gfield3}
\ee

\section {The matter coupling: vortex and the low-energy effective action }

The scalar  matter fields  $q$   in the fundamental representation  of $SU(N+1)$  (``squarks'') can be decomposed as 
\be   q^{a}_{I}(x) =   z^{a} \chi_{I} +  e^{a}_{i}\, \eta^{i}_{I}\;, \ee 
where 
\be   a=1,2, \ldots, N+1\;, \quad    i=1,2,\ldots, N\;,   \quad  I=1,2, \ldots N_{f}=N\;, \label{decompos}
\ee
namely,  into the component parallel to the symmetry breaking direction, $z$,  (see Eq.~(\ref{symbreak}))  and those orthogonal to it, $e^{a}_{i}$'s.

As one sees from the  fact that
\be   U^{\dagger} \, q_I=       U^{\dagger} (z \chi_I+ e \eta_I) =    \left(\begin{array}{c}\chi_I \\\eta^1_I \\\vdots \\\eta^N_I \end{array}\right)\;,
\ee 
$\chi$ and $\eta$'s are nothing but the  scalar field components  in the  singular gauge of the monopole (in which the adjoint scalar field does not wind and approaches a fixed VEV in all directions).
The monopole fields $N_{\mu}$, $P_{\mu}$ which couple to the projected scalars $\chi$ and $\eta$  have automatically the (asymptotic)  form of the   't Hooft-Polyakov monopole in the singular gauge.    As is well known the monopole field develops a Dirac string singularity attached to it in such a gauge.

By using the orthonormality and completeness of the vielbeins, one arrives at the following decomposition
\bea   \{(\de_{\mu} -  i  g  A_{\mu} )\, q_{I} \}^{a}&=&   z^{a} \{ \de_{\mu}\chi_{I} +  ({\bar z}\, \de_{\mu} z) \chi_{I} -  i g \frac{N}{N+1} \, C_{\mu}  \chi_{I} \} \nonumber \\
&+&  e^{a}_{j}\,\{  (\de_{\mu} -  i\,g \, b_{\mu})^{j}_{k}\, \eta^{k}_{I} +  {\bar e}^j_b\, \de_{\mu} e_i^b \, \eta_{I}^{i} +  i \frac{g}{N+1} \, C_{\mu}  \eta_{I}^{j} \}\;. 
\eea
Both $\chi$ and $\eta$ have  $U(1)$ electric charge, whereas  only  the $\eta$ fields carry non-Abelian $SU(N)$ charges.  
The matter kinetic term thus decomposes  as 
\bea   \Tr \,  |(\de_{\mu} -  i  g A_{\mu} )\, q_{I}|^{2} &=&  |    \de_{\mu}\chi_{I} - i g \frac{N_{\mu}}{2}   \chi_{I} -  i g \frac{N}{N+1} \, C_{\mu}
  \chi_{I} |^{2} \nonumber \\
&+&  | (\de_{\mu} -  i\, g\, b_{\mu})^{j}_{k}\, \eta^{k}_{I} - i g (P_{\mu})^{j}_{i} \, \eta_{I}^{i} +  i \frac{g}{N+1} \, C_{\mu}^{(0)}  \eta_{I}^{j}|^{2}\;.
\eea
The minimization of the potential Eq. (\ref{potential}) leads to 
\be   | ( \phi + m_{0} {\mathbbm 1} ) \, q
|^{2} =   |( m_{0} (N+1) M + m_{0} {\mathbbm 1} ) \, q|^{2} =  |\, m_{0}  (N+1)\, \chi_{I}  \,|^{2}  \;,\ee
which shows that the scalars in the $z$ direction, $\chi_{I}$,  are massive, 
which can be integrated out.   The low-energy effective action considered below describes the physics of the massless fields $\eta_{I}$, the "unbroken" gauge fields $b_{\mu}$  and the monopole  $ P_{\mu}$ \footnote{Unlike in some earlier attempts to study nonAbelian duality \cite{NAduality}  we here keep account only of the massless scalar and gauge degrees of freedom (besides the monopoles and antimonopole): they  are the relevant  degrees of freedom describing the infrared physics, i.e., at the mass scale below $v_1$.  The lower-mass-scale scale ($v_2$)  symmetry breaking and formation of the vortices are described by these light degrees of freedom.  The physics below the second symmetry breaking scale $v_2$, is the 
 massless Nambu-Goldstone like excitation modes living on the vortex-monopole world strip, the subject of our study below.},
\bea {\cal L} &=&   - \frac{1}{2} \Tr F_{\mu\nu}F^{\mu\nu} 
+
\sum_{I, j}\left| (\de_{\mu} -  i\, g\, b_{\mu})^{j}_{k}\, \eta^{k}_{I} - i g (P_{\mu})^{j}_{i} \, \eta_{I}^{i} +  i \frac{g}{N+1} \, C_{\mu}\,  \eta_{I}^{j}\right|^{2}
-  V_{\eta}\; \nonumber  \\
&\equiv & {\cal L}^{gauge} + {\cal L}^{scalar}  - V_{\eta}\;, 
\label{action}   \eea
where 
\bea  \Tr F_{\mu\nu}F^{\mu\nu} & = &   ( \frac{M_{\mu \nu}}{2}  +    \frac{N}{N+1} C_{\mu \nu} )^2 + \nonumber \\
&+&  \tr \,  \left( (K_{\mu \nu})^i_j   -  \frac{1}{N+1} C_{\mu \nu} \delta^i_j +   (b_{\mu \nu})^i_j  +   (h_{\mu \nu})^i_j  
 \right)^2   \label{gaugetensor} 
\eea
and   $M_{\mu \nu}, (K_{\mu \nu})^i_j, C_{\mu \nu}, (b_{\mu \nu})^i_j, (h_{\mu \nu})^i_j$ are defined in  Eqs.~(\ref{gfield1})-(\ref{gfield3}).
For the minimum monopole lying in the $(1, i+1)$ $su(2)$ subalgebra, (Eqs. (\ref{minimonopole}),  (\ref{minimonopole1})),     
 \be   {\bar z}\, \de_{\varphi} z  =    i \frac{1 - \cos \theta}{2}, \qquad 
   {\bar e}^i\, \de_{\varphi} e_i  =  -   {\bar z}\, \de_{\varphi} z\ = -  i \frac{1 - \cos \theta}{2},    \quad ({\rm no ~ sum ~ over ~}  i)\;,    \label{DiracWY}
  \ee
\be    {\bar e}^j\, \de_{\mu} e_i  = 0, \quad  j\ne i, \qquad         {\bar z}\, \de_{\theta} z  =   {\bar e}^i\, \de_{\theta} e_i  = 0\;. 
\ee
Eq.(\ref{DiracWY})  is precisely the Wu-Yang monopole in the singular gauge,  with Dirac string along the negative $z$ axis,   $z \in (-\infty, 0)$, 
\be    A_{\varphi}^{Dirac} =     \frac{1 - \cos \theta}{\rho  g} \;, \label{DiracMono}
\ee
where $\rho$ is the distance from the $z$ axis. Taking into account the factor $\tfrac{1}{2}$ due to  the $su(2) \subset su(N+1)$ embedding $\tfrac{\tau}{2}$,   
we see that the light scalars $\eta$ is coupled to such a monopole only through the $(P_{\mu})^{i}_{i}$ term: 
\be  P_{\mu}  = \frac{1}{2}
\left(\begin{array}{ccccc}\ddots &   &   &   &   \\  & 0 &   &   &   \\  &   & A_{\varphi}^{Dirac}  &   &   \\  &   &   & 0 &   \\  &   &   &   & \ddots\end{array}\right)  \label{Pmu}
\ee
The minimization of the potential  $V_{\eta}$   then leads to  the VEV (Appendix~\ref{minimum}) 
\be      \eta^i_{I}  =\delta^i_{I }  \,  {v_2} \;.  \label{VEVeta}
\ee
This VEV  brings the low-energy system into a color-flavor locked, completely Higgsed phase.  
In such a vacuum,  due to the exact flavor $SU(N)_{C+F}$ symmetry, broken by individual vortex solution, the latter develops non-Abelian orientational zero modes.

  \subsection{Symmetries  \label{sybsymm}}

The low-enegy system Eq.~(\ref{action}) arose from the gauge symmetry breaking, 
 \be    SU(N+1) \to   SU(N) \times U(1) 
 \ee
   and if the monopole fields $N_{\mu}$ and  $P_{\mu}$ are dropped    it would be the standard  $SU(N) \times U(1)$ gauge theory action.  
  It is clear, however,   that in the presence of a minimal monopole,  (Eqs. (\ref{minimonopole}),  (\ref{minimonopole1}),  (\ref{minimonopole2})),    the local color $SU(N)$ symmetry is broken by the  specific direction $M(x)= z {\bar z} -  (1/ (N+1)) {\mathbbm 1}_{N+1} $ the monopole points.  Attempts to define a global unbroken "orthogonal"  $SU(N)$ group in the presence of the 
 monopole background, lead inevitably to the well-known difficulties \cite{CDyons,DFHK}. 
 
 There are however some local and global symmetries which are left intact. In order to fix the idea, let us take the monopole in the  $(a,b)=  1,  2$  color subspace.  That is, we choose particular 
 monopole orientation with $i=1$, in Eqs. (\ref{minimonopole}),  (\ref{minimonopole1}),  (\ref{minimonopole2}).  In  this case the only nonvanishing component of  $P_{\mu}$ is: 
 \be   P^{1}_{\varphi\, 1}=\frac{i}{g} \,  {\bar e}^1\, \de_{\varphi} e_1  =  - \frac{i}{g}\,  {\bar z}\, \de_{\varphi} z\ = \frac{1 - \cos \theta}{2  g},    \qquad     P_{\mu \, i}^j=0,  \quad i\ne1, \, {\rm or} \,\, j\ne 1\;.      \label{monopol}\ee
 The action Eq.~(\ref{action}), Eq.~(\ref{gaugetensor}) is invariant under 
 \begin{description}
  \item[(i)]   a local $U(1)$  symmetry:
  \be    \eta^i \to   e^{i \alpha}  \eta^i, \qquad  C_{\mu}  \to  C_{\mu} -    \beta  \,  \de_{\mu} \alpha  \;, \qquad  \beta = (N+1)/g\;; 
  \ee
  
  \item[(ii)]  a local $U(1)$  symmetry: 
   \be    \eta \to    U \eta, \qquad b_{\mu}   \to      U   ( b_{\mu}  -  \frac{i}{g}  \de_{\mu }  )  U^{\dagger}\;,  \qquad      U=  \left(\begin{array}{cc}e^{i (N-1) \gamma} & 0 \\0 & e^{- i \gamma}\, {\mathbbm 1}_{N-1}\end{array}\right),    \ee 
    
  \item[(iii)] a local $SU(N-1)$ symmetry:
    \be    \eta \to     U   \eta, \qquad b_{\mu}   \to      U  ( b_{\mu}  - \frac{i}{g}  \de_{\mu }  )  U^{\dagger}\;,  \qquad     U=  \left(\begin{array}{cc}1 & 0 \\0 & V_{N-1}  \end{array}\right)\;, \ee 
\end{description}  
  (ii), (iii)  are subgroups of the local $SU(N)$  group broken by the particular orientation of the monopole.
  
 Finally, the action is invariant under    
   \begin{description}
  \item[(iv)] a global flavor $SU(N)_{F}$ symmetry: 
  \be    \eta \to  \eta \, {\cal U}^{\dagger},\qquad     {\cal U} \in  SU(N)\;.  \ee  
 \end{description}
  All the local ((i)-(iii)) and global ((iv)) symmetries are broken by the VEV of the scalars $\eta$,  Eq.~(\ref{VEVeta}).  However there remains 
  
  \begin{description}
  \item[(v)]  an exact global color-flavor diagonal $SU(N)$   symmetry 
 \be    \eta \to    {\cal U} \, \eta \, {\cal U}^{\dagger}, \qquad   P_{\mu}    \to   {\cal U}\, P_{\mu} \, {\cal U} ^{\dagger},      \qquad  b_{\mu} \to   {\cal U}\, b_{\mu} \, {\cal U}^{\dagger},  \label{CFsym}
\qquad    {\cal U} \in  SU(N)\;.
\ee
  Note that $K_{\mu \nu}, b_{\mu \nu},  h_{\mu \nu}$ all transform covariantly under Eq.~(\ref{CFsym}).  
 \end{description}
In particular, {\it the invariance of the action under (v)  requires that, together with the light matter and gauge fields, the monopole  $P_{\mu}= (i/g)\, {\bar e}\,\de_{\mu}e$  be also transformed with ${\cal U}$. } 

\subsection{Monopole-vortex soliton complex}  

  If it is not for the terms due to the monopole, $P_{\mu}$, $N_{\mu}$,  the action in  Eq.~(\ref{action})(Eq.~(\ref{gaugetensor})), would be exactly the $SU(N)\times U(1)$ gauge theory with  $SU(N)$ flavor symmetry where the vortices with nonAbelian $CP^{N-1}$ orientational zero modes has been first discovered \cite{Hanany:2003hp,ABEKY}.  The homotopy-group argument applied to the system with hierarchical gauge symmetry breaking (Eq. (\ref{SUNhierarchy})),  tells us that the vortex must end at the monopole. The nonAbelian orientational zero modes of the vortex endow  the endpoint monopoles with the same $CP^{N-1}$  zero modes.

  On the other hand, if the nonAbelian gauge fields  $b_{\mu}$ were neglected,  the above action would reduce to the low-energy $U(1)$ theory arising from the symmetry breaking of an $SU(2)$ gauge theory. The $SU(2)$ origin of such a theory is signaled by the presence of the monopole term:  performing the electromagnetic duality transformation explicitly \cite{Chatterjee}, keeping account of the presence of the monopole ($SU(2)/U(1)$ winding), one gets  an effective action of a static monopole acting as a source of the vortex emanating from it.   This analysis was repeated in the $\theta$ vacua of the original $SU(2)$ theory \cite{KMO}. 
 The resulting equation of motion has been solved analytically, reproducing the Witten effect correctly near the monopole  and showing a rather nontrivial  behavior of magnetic and electric fields near  the monopole-vortex complex. Our aim here is to generalize this construction  to a more general setting here, where both the vortex and monopole carry nonAbelian orientational zero modes. 
 
 The fact that the low-energy vortex must end at the monopole can be seen more directly.  The monopole term ${\bar e} \,\de_{\mu} e$  is really a non local term: 
 it contains the Dirac-string singularity running along the negative $z$-axis,  Eq. (\ref{DiracMono}).  In itself, it would give rise to an infinite energy, unless the scalar field vanishes precisely along the same half line $z \in  (-\infty, 0)$:  a vortex ending at the monopole  ($z=0$) and extending to its left.
 
\begin{figure}[ht]
\begin{center}
\includegraphics[width=0.8\linewidth]{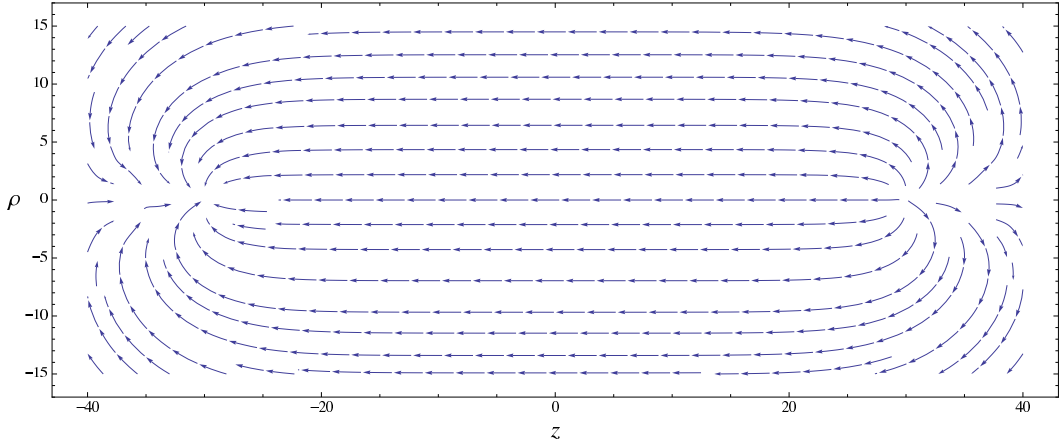}
\caption{\small  The magnetic field in the 
  monopole-vortex-antimonopole soliton complex. Taken from Cipriani, et. al.  \cite{CDGKM}}
\label{MattiaMVA}
\end{center}
\end{figure}

   A microscopic study of such a monopole-vortex complex has been made  by Cipriani et. al. \cite{CDGKM}, including the numerical determination of the field configurations 
    interpolating the regular 't Hooft-Polyakov monopoles to  the known vortex solution in between.  See Fig.~\ref{MattiaMVA} taken from \cite{CDGKM}.  We have not been successful so far  in generalizing  the derivation of the effective action for the orientational zeromodes directly  from the microscopic field-matter action as done for the nonAbelian vortex \cite{ABEKY,GJK}  to the present case of complex soliton of mixed codimensions.
   \footnote{Such a straightforward derivation of the effective action for the monopole-vortex mixed soliton systems has been achieved in \cite{CiprianiF}, in a particular BPS saturated model.  The model considered there  
   is different from ours: the ``monopole'' appears as a kink between the two degenerate 
   vortices, one Abelian and the other nonAbelian. 
    }  
   
\begin{figure}
\begin{center}
\includegraphics[width=4.5in]{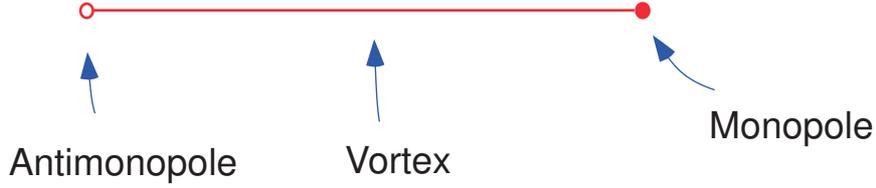}
\caption{\small The monopole, vortex and  anti-monopole complex of the preceding Figure, seen from large distances. }
\label{MVcomp}
\end{center}
\end{figure}

 Here we instead go to large distances first: the monopole is pointlike (this approximation has already been made),  and the vortex is a line, without width (see Fig. \ref{MVcomp}).
   Implementing this last approximation the scalar field takes the form,  
 \be   (\eta)^{i}_{I}  =     v_{2}\, \left(\begin{array}{cc}e^{i \psi}   & 0 \\0 & {\mathbbm 1}_{N-1} \end{array}\right)\;, \label{particular1}
 \ee
 whereas the relevant nonvanishing gauge fields are $C_{\mu}$,  and $(b_{\mu})^{i}_{j}$ of the form, 
 \be     (b_{\mu})^{i}_{j} =   \left(\begin{array}{cccc}(b_{\mu})^1_1 &   &   &   \\  & (b_{\mu})^2_2 &   &   \\  &   & \ddots &   \\  &   &   & (b_{\mu})^N_N\end{array}\right)\;.\label{particular2}
 \ee
 The monopole term is of the form, 
 \be      (P_{\mu})^{i}_{j} =   \left(\begin{array}{cccc}  - N_{\mu}/2 &   &   &   \\  & 0 &   &   \\  &   & \ddots &   \\  &   &   &   0 \end{array}\right)\;.\label{particular3}
 \ee
 Note the matched orientation in color for the vortex and monopole in Eqs. (\ref{particular1}) and (\ref{particular3}). 
 The scalar kinetic term in  the action, Eq. (\ref{action}), takes the form,  
 \be {\cal L}^{scalar} =   \sum_{I, j}   \left|  \left \{  \de_{\mu}  - i g \left(\begin{array}{ccc}(b_{\mu})^1_1 &   &      \\  & (b_{\mu})^2_2 &      \\  &   & \ddots   \end{array}\right) +  i g 
  \left(\begin{array}{ccc}  - N_{\mu}/2 &   &      \\  & 0 &      \\  &   & \ddots    \end{array}\right)  + \frac{i g \, C_{\mu}  }{N+1}  \, {\bf 1}_{N}  
\right \}^{j}_{i}  \, \eta_{I}^{i} \, \right|^{2}\;.      \label{particular0}
 \ee
 Clearly the minimum-energy condition for  $I=2,3, \ldots$ terms   requires that 
 \be    (b_{\mu})^{2}_{2}=  (b_{\mu})^{3}_{3}=\ldots =   (b_{\mu})^{N}_{N} =  \frac{1}{N+1} \,C_{\mu}\;.   \ee
 But as   $\tr \, (b_{\mu})  =0$,  this means that 
 \be      (b_{\mu})^{1}_{1} = -  (N-1)  (b_{\mu})^{2}_{2} =  - \frac{N-1}{N+1}\,C_{\mu}\;.   \label{particular4}
 \ee
 The $I=1$ term of Eq. (\ref{particular0})  then becomes  
  \bea {\cal L}^{scalar} &=& v_{2}^{2}\,  \left| \{  \de_{\mu} \psi - g  (b_{\mu})^{1}_{1}  + g   \frac{N_{\mu}}{2} + \frac{g  }{N+1}\, C_{\mu} \} \,\right|^{2} \nonumber \\
  &=&   (\de_{\mu}  \psi + g \frac{N_{\mu}}{2} +  g   \frac{N}{N+1}\,C_{\mu} )^{2}\, v_{2}^{2}\;.
 \eea
 On the other hand, the gauge kinetic term becomes
 \bea    {\cal L}^{gauge} &=&  -\frac{1}{2}  (\frac{M_{\mu \nu}}{2}  +  \frac{N}{N+1} C_{\mu \nu} )^{2} -  \frac{1}{2}  (\frac{M_{\mu \nu}}{2}   +  \frac{1}{N+1} C_{\mu \nu}   -   (b_{\mu \nu})^{1}_{1}  )^{2}  \nonumber \\
 &=&   -\frac{1}{4}   \, (M_{\mu \nu} +  \frac{2  N}{N+1} C_{\mu \nu} )^{2}\;. 
 \eea
 Far from the vortex-monopole, the potential can be set to be equal to its value in the bulk,  $V=0$.    The action reduces finally to the monopole-vortex Lagrangian 
 \bea  {\cal L}_{MV}   &=& 
  - \frac{1}{4}  (M_{\mu \nu} +   c_N \,  C_{\mu \nu})^{2}  +  (\de_{\mu}  \psi + g  \,  \frac{N_{\mu}}{2} +  \frac{g}{2}   \, c_N \,C_{\mu} )^{2}\, v_{2}^{2}\;, \qquad  
c_N   \equiv  \frac{2N}{N+1}\;  \nonumber \\
 &=&     - \frac{1}{4}  (  M_{\mu \nu} +   C_{\mu \nu} )^{2}  +  (\de_{\mu}  \psi + e  \,N_{\mu} +  e  \, C_{\mu} )^{2}\, v_{2}^{2}\;, 
     \label{reduced}
 \eea
 where in the last line we have re-normalized the Abelian gauge field $C_{\mu}$ by a constant and defined  the Abelian gauge coupling $e$   by  
 \be      c_{N}\, C_{\mu}  \to  C_{\mu}\;, \qquad    e \equiv  \frac{g}{2}. 
 \ee
In the discussion which follows, we take the  monopole fields  $N_{\mu}$ and $M_{\mu \nu}$ in Eq.~(\ref{reduced}) as a sum representing a monopole at ${\bf r}_{1}$ and an  anti-monopole at ${\bf r}_{2}$, at the two ends of the vortex.

\subsection{Orientational zeromodes }
  
  The action, Eq.~(\ref{action}), is invariant under the global color-flavor $SU(N)$ transformations, Eq.~(\ref{CFsym}).  The monopole-vortex field oriented in a particular direction, Eqs.~(\ref{particular1})- (\ref{particular4})  breaks this symmetry to $SU(N-1) \times U(1)$;  applying ${\cal U}$  on it
   \be    \eta \to    {\cal U} \, \eta \, {\cal U}^{\dagger}, \qquad   
     P_{\mu}    \to   {\cal U}\, P_{\mu} \, {\cal U} ^{\dagger},      \qquad  b_{\mu} \to   {\cal U}\, b_{\mu} \, {\cal U}^{\dagger},  \label{CFsymBis}
\ee
   generates a continuous set of degenerate configurations which span the coset, 
\be     \frac{SU(N)}{SU(N-1)\times U(1)} \sim CP^{N-1}.   
\ee
  The moduli space can be parametrized by 
 the so-called reducing matrix \cite{Delduc:1984sz},
  \begin{align}
{\cal U}(B)   =
\begin{pmatrix}
1 & - B^\dag \\
0 & \mathbf{1}_{N-1}
\end{pmatrix}
\begin{pmatrix}
X^{\frac{1}{2}} & 0 \\
0 & Y^{-\frac{1}{2}}
\end{pmatrix}
\begin{pmatrix}
1  & 0 \\
B & \mathbf{1}_{N-1}
\end{pmatrix}
=
\begin{pmatrix}
X^{-\frac{1}{2}} & - B^\dag Y^{-\frac{1}{2}} \\
B X^{-\frac{1}{2}} & Y^{-\frac{1}{2}}
\end{pmatrix}
\label{eq:Umatrix} \ ,
\end{align}
\beq
X\equiv 1  + B^\dag B \ , \quad
Y\equiv\mathbf{1}_{N-1}  + B B^\dag \ .   \label{Bmodes}
\eeq
acting on the light fields $\eta$ and $b_{\mu}$ and on the monopole field $e$, as in Eq.~(\ref{CFsym}).    $B$ is an $N-1$ component complex vector, the inhomogeneous coordinates of $CP^{N-1}$.

The  fields   corresponding to the particular  ``$(1,1)$''  orientation of the vortex-monopole,  Eqs.~(\ref{particular1})-(\ref{reduced}),    are of the form, 
\bea  && \eta=  e^{ i \psi}\, \frac{{\mathbbm 1}_{N}+ T }{2} +   \frac{{\mathbbm 1}_{N}-  T }{2}, \qquad 
\de_{\mu}  \eta   =  \de_{\mu} e^{ i \psi}\,  \frac{{\mathbbm 1}_{N}+ T }{2}\;,  
\nonumber \\
  &&  - b_{\mu} +  \frac{1}{N}  C_{\mu}  {\mathbbm 1}_{N} =   \frac{N}{N+1}   \frac { {\mathbbm 1}_{N} +T}{2}  \, C_{\mu}, \nonumber \\
&&   P_{\mu}=  - N_{\mu} \,  \frac{  {\mathbbm 1}_{N} +T }{2}, \qquad     T \equiv   \left(\begin{array}{cc}1 &   \\  & - {\mathbbm 1}_{N-1}\end{array}\right)\;. \label{pform}
\eea
The action is then calculated to be
\be       \tr   \left(\frac{{\mathbbm 1}_{N}+ T }{2} \right)^{2}  \cdot  {\cal L}_{MV}    =   {\cal L}_{MV} \;
\ee
where ${\cal L}_{MV}$ is given in Eq.~(\ref{reduced}).
When a global ($x_{\mu}$-independent)  color-flavor transformation  ${\cal U}$ acts on it, a new solution is generated by   
$   \{ \eta, b_{\mu}, P_{\mu}\}   \to  {\cal U} \{ \eta, b_{\mu}, P_{\mu}\}  {\cal U}^{\dagger}\;,  $  that is
\bea  && \eta=  e^{ i \psi}\, \frac{{\mathbbm 1}_{N}+  {\cal U} T  {\cal U}^{\dagger} }{2} +   \frac{{\mathbbm 1}_{N}-   {\cal U} T  {\cal U}^{\dagger}}{2}, \qquad 
\de_{\mu}  \eta   =  \de_{\mu} e^{ i \psi}\,  \frac{{\mathbbm 1}_{N}+  {\cal U} T  {\cal U}^{\dagger}}{2}\;,  
\nonumber \\
  &&  - b_{\mu} +  \frac{1}{N}  C_{\mu}  {\mathbbm 1}_{N} =   \frac{N}{N+1}   \frac { {\mathbbm 1}_{N} + {\cal U} T  {\cal U}^{\dagger}}{2}  \, C_{\mu}, \nonumber \\
&&   P_{\mu}=  - N_{\mu} \,  \frac{  {\mathbbm 1}_{N} + {\cal U} T  {\cal U}^{\dagger}}{2}\;.  \label{rotform}
\eea
All the fields now have complicated, nondiagonal forms  both in color and flavor spaces.   
Note however  that 
\be   \Pi_B\equiv  \frac{{\mathbbm 1}_{N}+ {\cal U} T  {\cal U}^{\dagger} }{2}\;,  \qquad  \Pi_{O}\equiv  \frac{{\mathbbm 1}_{N}- {\cal U} T  {\cal U}^{\dagger} }{2}\;, 
\ee
act as the projection operators to the directions in color-flavor space, along the vortex-monopole orientation and perpendicular to it:
\be        \Pi_B^{2}=   \Pi_B, \quad   \Pi_{O}^{2}=   \Pi_{O},    \quad   \Pi_B\cdot  \Pi_{O} = 0\;;   \qquad   \tr \, \Pi_B^{2}=   1\;.
\ee
By using these,  the action corresponding to the color-flavor rotated configuration, Eq.~(\ref{rotform})   is seen to be still given by 
\be    \tr\,  {\cal L}_{scalar}  \, \Pi_B^{2}   =   {\cal L}_{scalar}\,,     \qquad  \tr\,  {\cal L}_{gauge}   \, \Pi_B^{2}    =   {\cal L}_{gauge}\,,     \qquad   \tr\,  \Pi_B^{2}  \,  {\cal L}_{MV}    =   {\cal L}_{MV} \;,
\ee
reflecting the exact $CP^{N-1}$ moduli of the monopole-vortex solutions,   following from the breaking of the exact color-flavor   symmetry, Eq.~(\ref{CFsym}). 
Therefore the $CP^{N-1}$  modes $B$ of Eqs.~(\ref{CFsymBis})-(\ref{Bmodes})  represent exact zero modes of the monopole-vortex action, Eq.~(\ref{action}).

\subsection{Spacetime dependent $B$  \label{STdepB}}

The configurations Eq.~(\ref{particular0}) - Eq.~(\ref{reduced}), or the color-flavor rotated version, Eq.~(\ref{rotform}),  represents the long-distance approximation of the 
nonAbelian vortex with monopoles attached at the ends.  They are basically an  Abelian configuration {\it embedded} in a particular direction in $SU(N)_{C+F}$, 
  \[ \Pi_B\equiv  \frac{{\mathbbm 1}_{N}+ {\cal U}(B)  T  {\cal U}(B)^{\dagger} }{2}\;, 
\]
This is so (i.e., Abelian)  even if the scalar field and gauge (and monopole) fields all have nontrivial matrix form in general in color and flavor, as they all commute with each other. 

\begin{figure}
\begin{center}
\includegraphics[width=4in]{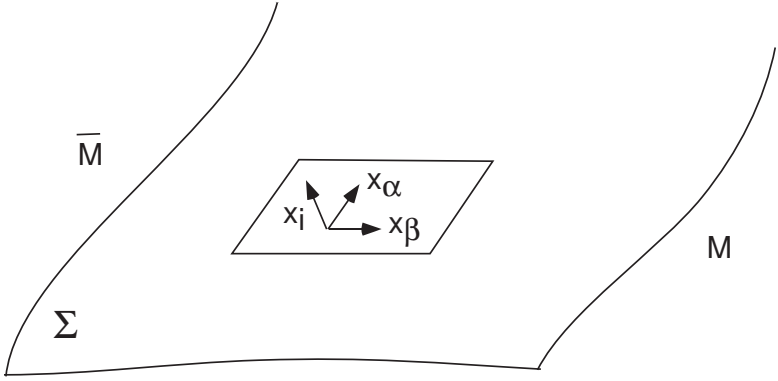}
\caption{ \small   Worldsheet strip  $\Sigma$  spanned between the worldlines of the monopole and antimonopole. }
\label{Worldsh}
\end{center}
\end{figure}

When the orientational moduli parameter $B$ is made to depend weakly on the spacetime variables $x_{\mu}$, however,  such an Abelian structure cannot be maintained. 
The derivative acting on $\Pi_B$ in the scalar field  induces the change of charge and current 
 \[   \de_{\mu}  (\eta \Pi_B) =    (\de_{\mu} \eta )   \Pi_B +   \eta \, \de_{\mu}  ( {\cal U}  T  {\cal U}^{\dagger})
\]
along the vortex.  It implies, through the equations of motion,
 \be
   \frac{1}{g^2} {\cal D}^{i}  F_{i \alpha}^a    =    i\,  \sum_{I} \big[ \eta_{I}^{\dag} \,t^a {\cal D}_{\alpha}
      \eta_{I} - ({\cal D}_{\alpha} \eta_{I})^{\dag} t^a  \eta_{I} \big], 
\label{BioSavart}
\ee
new gauge field components, $A_{\alpha}^{(B)}$.  This can be understood as  nonAbelian  
Biot-Savart or Gauss' law.     Following  \cite{GJK}   we have introduced the index $\alpha$  to indicate the two spacetime coordinates in the vortex-monopole worldsheet $\Sigma$, while  indicating with  $``i"$ the other two coordinates of the plane perpendicular  to the vortex length.   See Fig.~\ref{Worldsh}.    For a straight vortex  in the $\hat z$ direction,  $\alpha=3,0$ whereas $i=1,2$.  By assumption $B$,  hence  ${\cal U}$, is a slowly varying function of $x_{\alpha}$.    It is not difficult to show  \footnote{ $A_{i}$ and $\sum_{I} \eta_{I} \eta_{I}^{\dagger}$  have both the form 
$a_{1} {\mathbbm 1} +  a_{2} \,{\cal U} T {\cal U}^{\dagger}$, where $a_{1,2}$ are some functions of the transverse variables $x_{i}$.    $\de_{\alpha}$ acts only on ${\cal U}$.  Repeated use of   
$$\de_{\alpha} ({\cal U} T {\cal U}^{\dagger}) \, {\cal U} T {\cal U}^{\dagger}  \, {\cal U} T {\cal U}^{\dagger} = \de_{\alpha} ({\cal U} T {\cal U}^{\dagger}),
  \qquad [\de_{\alpha} ({\cal U} T {\cal U}^{\dagger})  {\cal U} T {\cal U}^{\dagger},     {\cal U} T {\cal U}^{\dagger} ] = 2\,  \de_{\alpha} ({\cal U} T {\cal U}^{\dagger})$$   and  $ ({\cal U} T {\cal U}^{\dagger})^{2} = {\mathbbm 1}$   in Eq.~(\ref{BioSavart}) yields  Eq.~(\ref{Gauss}). } that  $A_{\alpha}^{(B)}$ is  oriented in the direction
\be      A_{\alpha}^{(B)} \propto  \de_{\alpha} (  {\cal U}  T  {\cal U}^{\dagger}) \,  {\cal U}  T  {\cal U}^{\dagger}  =  2\,   {\cal U}  (   {\cal U}^{\dagger} \de_{\alpha}  {\cal U} )_{\perp} \,  {\cal U}^{\dagger},   \label{Gauss}
\ee
in color-flavor mixed space,   where 
\be   ({\cal U}^{\dagger} \de_{\alpha}  {\cal U} )_{\perp} =  \frac{1}{2}( {\cal U}^{\dagger} \de_{\alpha}  {\cal U}  - T \,{\cal U}^{\dagger} \de_{\alpha}  {\cal U} \,T )\;.  
\ee
 $ ({\cal U}^{\dagger} \de_{\alpha}  {\cal U} )_{\perp} $ is just the Nambu-Goldstone modes  \cite{Delduc:1984sz,GJK}  in a fixed 
vortex background, Eq.~(\ref{pform});    as the vortex-monopole rotates  (\ref{rotform}), one has to rotate them in order to  keep them orthogonal to the latter.

The effect is to produce the  electric and magnetic fields lying in the plane perpendicular to the vortex direction, $F_{i \alpha}$,  along the vortex.  

By using the orthogonality relations
\be    \tr \, \Pi_B\, \de_{\alpha} (  {\cal U}  T  {\cal U}^{\dagger})  = \tr\, \Pi_B\,   \de_{\alpha} (  {\cal U}  T  {\cal U}^{\dagger}) \,  {\cal U}  T  {\cal U}^{\dagger}   =0,    \label{orto}
\ee
it is easily seen that the terms containing the derivatives $\de_{\mu} {\cal U}$ or $\de_{\mu}{\cal U}^{\dagger}$ give rise to the correction  
 \bea  &&  {\cal L} ({\{ \eta, b_{\mu}, P_{\mu}\}   \to  {\cal L} ({\cal U} \{ \eta, b_{\mu}, P_{\mu}\}  {\cal U}^{\dagger}} )  =   {\cal L}_{scalar} + \delta {\cal L}\;,  
\nonumber \\
 &&  \Delta {\cal L}  ={\rm const.}  \,   \tr  \left( \de_{\alpha} ({\cal U} T  {\cal U}^{\dagger})  \right)^2     \propto  \tr\left\{
X^{-1} \p_\alpha B^\dag
Y^{-1}  \p_\alpha B
\right\} \ , 
\eea
which yields  the well-known $CP^{N-1}$ action  
  \be
S_{1+1}   
=  2\beta \int_{\Sigma} d^{2}x\; \tr\left\{
\left(\mathbf{1}_N + B^\dag B\right)^{-1}\p_\alpha B^\dag
\left(\mathbf{1}_N + B B^\dag\right)^{-1}\p_\alpha B
\right\} \ , \label{eq:sigmamodelaction}
\ee
where the coupling constant $\beta$  arises as the result of integration of the vortex-monopole profile functions, in the plane perpendicular to the vortex axis. 
The $x_{\alpha}$-dependence through $B$, by definition at much larger wavelengths than the vortex width / monopole size, factorizes and give rise to the 
$CP^{N-1}$  action defined on the worldstrip.

A proper derivation of such a $2D$ worldsheet action for the vortex system including the determination of $\beta$   requires to maintain a profile functions $f(r)$ in (\ref{particular1}),
$ e^{i \psi}  \to  f(r)\, e^{i \psi} $,  $f(0)=0; \, f({\infty})=1$  and study its equation of motion. Although this can be done straightforwardly for the pure vortex configuration (without  monopoles) \cite{Klee, ohashi}, the analysis has not been done in the presence of the endpoint monopoles. 
We plan to come back to this more careful analysis elsewhere. 
A microscopic study of the vortex in a non-BPS system which is very close to our model, has been done by Auzzi et. al.  \cite{AEV}, without however attempts to determine the vortex effective action.

 \section{Dual description}
 
 Following \cite{Nambu,Klee,Akhmedov,Chatterjee,KMO} we now dualize the system, Eq.~(\ref{reduced}):  
  \be    - \frac{1}{4}  (M_{\mu \nu} +   C_{\mu \nu} )^{2}  +  (\de_{\mu}  \psi + e  \,N_{\mu} +  e  \, C_{\mu} )^{2}\, v_{2}^{2}\;.  \label{reducedBis}
 \ee 
 All the fields above  live in the particular direction in the color-flavor space, for instance, 
 \be       \Pi_B^{(0)}\equiv  \frac{{\mathbbm 1}_{N}+ T }{2} =   \left(\begin{array}{cccc}1 & 0 & \hdots & 0 \\0 & 0 &  & 0 \\ &  & \ddots & \vdots \\0 & 0 &  & 0\end{array}\right)\;,
 \ee
 (see Eq.~(\ref{particular1})-Eq.~(\ref{particular3})).     The factor $\tr (\Pi_B^{(0)})^{2} =1$ in the action is left implicit. 
  Decompose $\psi$ field into its regular and singular part:
 \be   \psi = \psi^{r}  + \psi^{s}\;. 
 \ee
 The latter (non-trivial winding of the scalar field) is related to the vortex worldsheet  loci  by \cite{Nambu,Klee,Akhmedov}
 \beq  &&  \epsilon^{\mu \nu \rho\sigma}\de_{\rho} \de_{\sigma} \psi^{s} \equiv   \Sigma^{\mu \nu}(x)   \nonumber \\
 &=& 2\pi n \, \int_{\Sigma} \de_{a}  x^{\mu} \de_{b}x^{\nu}  
(d\xi^{a} \wedge d \xi^{b})\, \delta^{4}(x-x(\xi)) \;
  \label{worldsheet}  \eeq 
and $\xi^{a}= (\tau, \sigma)$, $\sigma \in (0, \pi)$,   are the worldsheet coordinates and $n$ is the winding.   $\Sigma^{\mu \nu} $ is often referred to as the vorticity in the literature.   Below  we shall limit  ourselves to the case $n=1$  (the minimum winding)  for the purpose of studying  the 
transformation properties of the vortex and monopole
\footnote{   Eq.~(\ref{worldsheet}) can be seen as the change of field variables from $\psi (x)$ to the string variable $x_{\mu}(\tau, \sigma)$.    Keeping track of the Jacobian of this transformation leads to the Nambu-Goto action,   $\int  d\tau d\sigma \,(\det|\de_{a}  x^{\mu} \de_{b} x^{\nu}|)^{1/2}$, describing the string dynamics,  and possible corrections.
We shall not write these terms explicitly below in the effective action  as  our main interest lies in the internal, color flavor, orientational zeromodes.}.    
 We assume that the monopole and anti-monopole are at the edges of the worldstrip  ($\sigma=0, \pi$):
\be   {\bf r}_{1}= {\bf r}(\tau, 0), \qquad   {\bf r}_{2}= {\bf r}(\tau, \pi).    \label{monopoles}\ee
It then follows from Eq.~(\ref{worldsheet}) that 
\be   \de_{\mu}    \Sigma^{\mu \nu}(x) =  2\pi   j^{\nu}   \label{builtin}
\ee
where $j^{\nu}$ represents  the  monopole and antimonopole currents:
\be    j^{\nu}  =  \int d\tau  \frac{d x^{\nu}}{d \tau}  \, \delta^{4}( x -  x(\tau, \pi)) -    \int d\tau  \frac{d x^{\nu}}{d \tau}  \, \delta^{4}( x -  x(\tau, 0)),    \label{mechcurrent}
\ee    
with   $ x^{\mu}(\tau, \pi)$ and  $x^{\mu}(\tau, 0)$ standing for their worldlines.  We shall see below that the equations of the dual system consistently reproduces this ``monopole confinement''
condition.

The regular part $\psi^{r}$ can be integrated out by introducing the Lagrange multiplier 
\beq  - \frac{1}{4  v_{2}^{2}} \lambda_{\mu}^{2}  +  \lambda_{\mu} \,   ( \de_{\mu} \psi^{r} + \de_{\mu} \psi^{s} +  e N_{\mu} + e\,C_{\mu} )\;,   
\eeq
which gives rise to a functional delta function
\beq   \delta (\de_{\mu}   \lambda_{\mu}(x))\;.
\eeq
The constraint can be solved by introducing an antisymmetric field  $B_{\mu \nu}(x)$, 
\beq    \lambda^{\mu} =  \frac{v_{2}}{ \sqrt{2}} \,    \epsilon^{\mu \nu \rho \sigma}   \de_{\nu} B_{\rho \sigma} =   \frac{v_{2}}{3 \sqrt{2}}    \epsilon^{\mu \nu \rho \sigma} H_{\nu \rho \sigma }\;, 
\eeq
\beq 
H_{\nu \rho \sigma } \equiv   \de_{\nu} B_{\rho \sigma} +  \de_{\rho} B_{\sigma \nu} +  \de_{\sigma} B_{\nu \rho}  \nonumber
\eeq
   being a completely antisymmetric tensor field.  One is left with the 
Lagrangian 
\beq  {\cal L} &=&   - \frac{1}{4}  (M_{\mu \nu} +  \, C_{\mu \nu} )^{2}  + \, 
\frac{ e \,  v_2  \, }{\sqrt{2}} \,   \epsilon^{\mu \nu \rho \sigma}  C_{\mu} \de_{\nu} B_{\rho \sigma}   \nonumber \\
  &+&   \frac{1}{12}  H_{\mu \nu \lambda}^{2}  +\frac{v_2}{ \sqrt 2 }   \, B_{\mu \nu} \Sigma^{\mu \nu} +  \frac{e \, v_2}{2\sqrt 2}  \epsilon^{\mu \nu \rho \sigma}   M_{\mu \nu}  B_{\rho \sigma}\;\;.
\eeq 
Now we dualize  $C_{\mu}$ by writing \footnote{This is the standard Legendre transformation of the electromagnetic duality. }
\bea  &&  \int [d C_{\mu}]    \exp{i \int d^{4}x \,  \{  - \frac{1}{4}  (M_{\mu \nu} +  \, C_{\mu \nu} )^{2}  + \, 
\frac{e \,v_2 \, }{\sqrt{2}} \,   \epsilon^{\mu \nu \rho \sigma}  C_{\mu} \de_{\nu} B_{\rho \sigma} \}} \nonumber \\
&& =    \int \, [d C_{\mu}]  [d \chi_{\mu \nu}]  \,  \exp{i \int d^{4}x \,  \{  -   \chi_{\mu \nu}^2  +     \chi_{\mu \nu} \, \epsilon^{\mu \nu \rho \sigma} (M_{\rho \sigma} +  \, C_{\rho \sigma} ) /2    + \frac{e\, v_2\,}{\sqrt{2}} \,   \epsilon^{\mu \nu \rho \sigma}  C_{\mu} \de_{\nu} B_{\rho \sigma}     \} }  \nonumber \\
&& =   \int    [d \chi_{\mu \nu}]   \,  \delta ( \epsilon^{\mu \nu \rho \sigma}    \de_{\nu}    (\chi_{\rho \sigma} +   e \, v_2 \, B_{\rho \sigma}/\sqrt{2} ))  
 \exp{i \int d^{4}x \,  \{  -   \chi_{\mu \nu}^2  +     \chi_{\mu \nu} \epsilon^{\mu \nu \rho \sigma} M_{\rho \sigma}/2   \} }\;.
\label{from}\eea
Again the constraint can be solved by setting 
\beq     \chi_{\mu \nu} =    \frac{1}{\sqrt{2} }(  \de_{\mu}   A_{D\,  \nu} - \de_{\nu}   A_{D\,  \mu}  -  \sqrt{2} \, e\,v_2  \,  B_{\mu \nu} ) 
\label{andfrom}  \eeq
and taking the dual gauge field $A_{D\, \mu}$ as the independent variables. 
 The Lagrangian is now
\beq  {\cal L}=   \frac{1}{12}  H_{\mu \nu \lambda}^{2}    - \frac{1}{4}(  \de_{\mu}   A_{D\,  \nu} - \de_{\nu}   A_{D\,  \mu} - \sqrt{2} \,e\, v_2  \, B_{\mu \nu})^{2}  
+ \frac{v_2}{ \sqrt 2 }   \, B_{\mu \nu} \Sigma^{\mu \nu} +    A_{D \mu } \, J^{\mu}\;, \label{integrateAD}
\eeq 
where
\be  H_{\nu \rho \sigma } \equiv   \de_{\nu} G_{\rho \sigma} +  \de_{\rho} G_{\sigma \nu} +  \de_{\sigma} G_{\nu \rho}\;,  \qquad 
G^{\mu \nu}\equiv    B^{\mu \nu} - \frac{1}{\sqrt{2} \, e\, v_2 }( \de^{\mu} A_{D}^{\nu} -  \de^{\nu} A_{D}^{\mu}) \;,    \label{Hmn}\ee
and 
\beq      J^{\mu} =  \de_{\nu} \, \frac{1}{2} \epsilon^{\mu \nu \rho \sigma}  M_{\rho \sigma}  =   \de_{\nu} \, {\tilde   M}^{\mu \nu}  
\eeq 
represents the monopole magnetic current \footnote{To distinguish the monopole magnetic charge current from the point-particle ``mechanical'' current  (Eq.~(\ref{builtin})), we use $J^{\mu}$ (for the former) and $j^{\mu}$ (for the latter), respectively.   See Eq.~(\ref{constraintnew}) below.  }.   One sees from Eq.~(\ref{from}) and Eq.~(\ref{andfrom})  that $A_{D}^{\mu}$ is indeed   locally coupled to $ J^{\mu} $. 
Finally, observing that there is a (super) gauge invariance of the form,  
\beq  \delta  B_{\mu \nu} = \frac{1}{\sqrt{2} \, e\, v_2}  ( \de_{\mu} \Lambda_{\nu} -  \de_{\nu} \Lambda_{\mu});\qquad  \delta A_{D}^{\mu}=   \, \Lambda^{\mu}\;, \label{ADgauge}  \eeq
one  can write the Lagrangian in terms of the gauge-invariant field $G_{\mu \nu}$, 
\beq  {\cal L}=   \frac{1}{12}  H_{\mu \nu \lambda}^{2}    - \frac{m^{2}}{2} \, G_{\mu \nu}^{2} +     \frac{v_2}{ \sqrt 2 }  \, G_{\mu \nu}   \Sigma^{\mu \nu}\;, \qquad m \equiv  e \, v_2 \;. \label{final}
\eeq
Note that use of the gauge invariance under, Eq.~(\ref{ADgauge}) - or the integration over $A_D$ in Eq.~(\ref{integrateAD})  -   introduces a constraint
\beq        \de_{\mu}\, \Sigma^{\mu \nu} =  e   \,  J^{\nu}.  \label{constraint}
\eeq

Let us comment on the relation between this equation and  the constraint, (\ref{builtin}), (\ref{mechcurrent}).   For the static  minimum monopole, Eq.~(\ref{gfield1}), with the form of $z$ given in  Eq.~(\ref{minimonopole}),  
one finds  
\be  J^{0} =  \de_{\nu} \, \frac{1}{2} \epsilon^{0  \nu \rho \sigma}  M_{\rho \sigma}  =   \de_{i} \, \frac{1}{2} \epsilon^{ijk }  M_{jk} =  \de_i  B_i, 
\ee
where $B_i$ is the  magnetic Coulomb  field,  
\be    B_i =  -  \frac{1}{g}  \nabla_i  \frac{1}{r}     \label{mfield}
 \ee
 following from Eq.~(\ref{gfield1}) and  Eq.~(\ref{DiracWY}) \footnote{We  recall that the form of the vector potential Eq. (\ref{DiracWY})  in the cylindrical coordinates gives precisely the isotropic 
 Coulomb magnetic field. }.   Thus 
 \be  J^{0} =   - \frac{1}{ g}   \nabla \cdot \nabla  \frac{1}{r}  =  \frac{4\pi}{g}   \delta^3 ({\bf r})       
 \ee
 showing that it has the well-known magnetic charge, 
 \be    g_m =    \frac{4  \pi}{g}\;, 
 \ee
 of a 't Hooft-Polyakov monopole, consistent with the Dirac condition  (after setting  $g= 2 \, e$). 
 Equation (\ref{constraint})  then becomes 
 \beq        \de_{\mu}\, \Sigma^{\mu 0 } =  e   \,  J^{0} = 2 \pi  j^0,    \label{constraintnew}
\eeq
 where $j^0(x)=    \delta^3({\bf r})$ is the mechanical particle current.  This is indeed the monopole confinement condition,    Eq.~(\ref{builtin}) and Eq.~(\ref{mechcurrent}).

Inverting the logic, one may say that the (super) gauge invariance Eq.~(\ref{ADgauge}) hence the possibility of writing the effective action in terms of the gauge field $G_{\mu \nu}$, 
{\it follows from}  the built-in  monopole confinement condition,   $\de_{\mu}  \Sigma^{\mu \nu}= 2\pi   j^{\nu}$.

The monopole thus  acts as the  source (or the sink) of the worldsheet  and at the same time plays the role of  the magnetic point source for the dual gauge fields. 
The equations of motion for $G_{\mu \nu}$  are:
\beq    \de_{\lambda}  H^{\lambda \mu \nu} =  - m^{2} \,   G^{\mu \nu} +  \frac{ m}{\sqrt{2} \,e} 
\, \Sigma^{\mu \nu}\;.   \label{dualeq}   \eeq
By taking a further derivative and by using Eq.~(\ref{constraint})    one finds 
\beq  \de_{\mu}   G^{\mu \nu} = \frac{
1}{ \sqrt{2}\, m}  \, J^{\nu}\;.  \label{eqmdual}
\eeq 

These equations of motion have been studied  in \cite{KMO}  in the general case of a $\theta$ vacuum of the original high-energy theory.   The main results are reported in Appendix~\ref{metric}.     There are a few remarkable features which will  be useful below.  First of all, outside the worldsheet strip,  $\Sigma_{\mu \nu}=0$, so Eq.~(\ref{dualeq}) tells  that  $G_{\mu \nu}$ are massive field which die out exponentially in all directions, away from the monopole-vortex complex. Second, there are two distinct nonvanishing components  of $G_{\mu \nu}$.  One  is what makes up the 
vortex cloud, the dominant part being the constant (in the vortex length direction)  magnetic field along the vortex direction, and having a transverse thickness of the order of $1/gv_{2}.$  Another component is a spherically symmetric, Coulomb magnetic field cloud around the monopole, of the radial size, $1/g  v_{1}$.   This includes, in the $\theta$ vacuum, the Coulomb electric field due to the Witten effect.

\section{Orientational $CP^{N-1}$ zeromodes in the dual theory}

The dualization procedure above can be repeated starting from the monopole-vortex complex of generic orientation, Eq.~(\ref{rotform}).  The result is the effective action having the identical form as Eq.~(\ref{final})  but with all fields replaced by 
\be   G_{\mu \nu}^{(0)} \to   G_{\mu \nu}^{(B)} \equiv    G_{\mu \nu}^{(0)}  \Pi_B \;, \qquad      \Pi_B \equiv    \frac{{\mathbbm 1}_{N}+ {\cal U}(B)  \, T  \, {\cal U}^{\dagger}(B) }{2}\;; 
\label{replace1}\ee
\be     H_{\mu \nu \lambda}^{(0)} \to     H_{\mu \nu \lambda}^{(B)} \equiv   \de_{\mu} G_{\nu \lambda}^{(B)}  +  \de_{\nu} G_{\lambda \mu}^{(B)}  +  \de_{\lambda} G_{\mu \nu}^{(B)}  \;;
\ee
\be   J^{\mu} =  \de_{\nu} \, \frac{1}{2} \epsilon^{\mu \nu \rho \sigma}  M_{\rho \sigma}  =   \de_{\nu} {\tilde   M}^{\mu \nu}    \to   \de_{\nu} ({\tilde   M}^{\mu \nu} \Pi_B )= 
 \de_{\nu} {\tilde   M}^{\mu \nu\, (B)}\;.
\ee
The equation of motion for $H^{\sigma \mu \nu}$
\beq    \de_{\lambda}  H^{\lambda \mu \nu  \,  (B)}  =  - m^{2} \,   G^{\mu \nu   \, (B)}  +  \frac{ m}{ \sqrt{2}\, e} 
\, \Sigma^{\mu \nu \,  (B)} \;,   \qquad   m\equiv  g v_{2},  \label{eqmdual1}\eeq
and the equation for  $G^{\mu \nu} $ (which follows by differentiating the above and using Eq.~(\ref{builtin})) 
\beq  \de_{\mu}   G^{\mu \nu \,   (B)}  = \frac{1}{\sqrt{2} \, m}  \,  J^{\nu \, (B)} \;  \label{eqmdual2}
\eeq
all  hold with  $G^{\mu \nu \,   (B)}$ and  $ J^{\nu \, (B)}$  lying in the color-flavor direction $\Pi_B$.     As 
\be     \tr \, \Pi_{B}^{2}  =1,
\ee
the action 
\beq  {\cal L}=   \frac{1}{12}  \tr\, (H_{\mu \nu \lambda}^{(B) \,  2} )   - \frac{m^{2}}{4} \, \tr \,( G_{\mu \nu}^{(B) \,  2} ) +     \frac{1}{ \sqrt 2 }  \,\tr\, (  G_{\mu \nu}^{(B)}   \Sigma^{\mu \nu \, (B)}) \;  \label{finalAction}
\eeq
is independent of $B$, as long as $B$ is  constant .

\begin{figure}
\begin{center}
\includegraphics[width=5in]{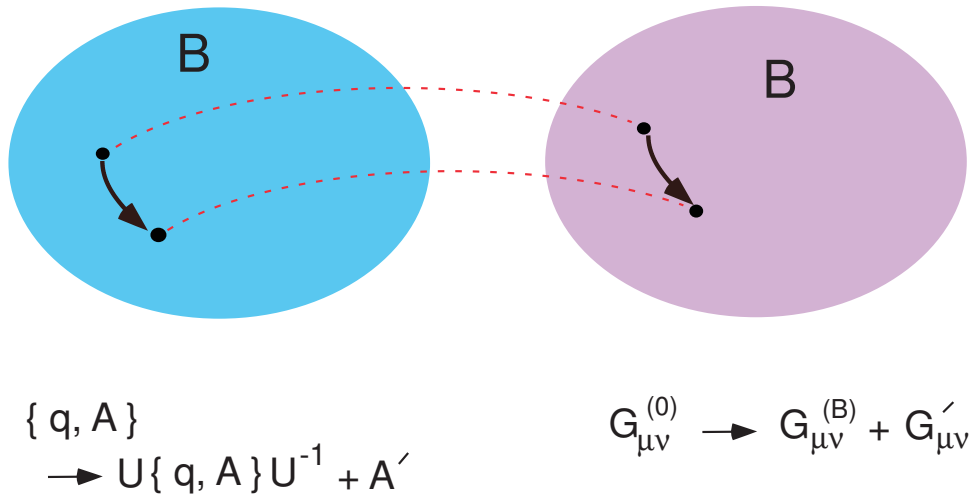}
\caption{\small The muduli space of monopole-vortex configuration in the electric variables (the left figure) and in the magnetic dual variables (on the right).  The  points in the two $CP^{N-1}$ spaces are in one-to-one correspondence.   The motion in $CP^{N-1}$ moduli space  in real space-time requires the fields to be transformed nontrivially, both in the electric and magnetic descriptions. }
\label{onetoone}
\end{center}\end{figure}

We find therefore a $CP^{N-1}$ moduli space of degenerate monopole-vortex configurations described by the dual variables, Eq.~(\ref{finalAction}),  
each point of which is in one-to-one correspondence with that of the the monopole-vortex solutions in the electric description,  Eq.~(\ref{rotform}). 
See Fig.~\ref{onetoone}. 

\subsection{Spacetime dependent  $B$ and the effective action}

Allow now the orientational moduli parameter $B$ to fluctuate in spacetime.
A na\"{i}ve guess is  to assume 
\be G_{\mu \nu}^{(0)} \to   G_{\mu \nu}^{(B)}   \label{simple}\ee
 also for spacetime dependent $B$ and to study  the excitation of the action due to the derivatives $\de / \de x_{\alpha}$  acting on  $\Pi_B$ in the kinetic term of $G_{\mu \nu}^{(B)}$, $\frac{1}{12}  \tr\, (H_{\mu \nu \lambda}^{(B)})^{2},$  i.e.,   
$H_{\mu \nu \lambda}^{(B)} \equiv   \de_{\mu}  G_{\nu \lambda}^{(B)} +  ({\rm cyclic})$.
This however is not correct. 

In moving  in the $CP^{N-1}$  moduli space one  must make sure that the massive modes are not excited.  
This is the reason, in the electric description,  for the introduction of the new gauge modes $A_{\alpha}$,  such that the YM-matter equations of motions continue to be satisfied, i.e.,  to  stay in the minimum-action subspace.  In other words, one must stay on the minimum  subsurface \footnote{As we have noted already, this is a nonAbelian analogue of Gauss or Biot-Savart law.
In a closer context of soliton physics, this is also the essence of Manton's ``moduli-space approximation''  \cite{Manton}  for describing the slow motion of soliton monopoles, although here we are  concerned with the "motion" of the soliton monopole-vortex in the internal color-flavor space .
} 
\be  \frac{\delta S}{\delta G_{\mu \nu}}  =0 \label{minin}
\ee
in all (color-flavor) directions.  For $\delta G_{\mu \nu} \propto \Pi_{B}$, this is automatic;   this condition must also satisfied for orthogonal fluctuations  $\delta G_{\mu \nu} \propto \de_{\alpha} \Pi_{B}$, if the  system tends to generate  them.  
Quadratic fluctuation in ${\de_{\alpha}B}$  computed at such minimum trough then gives the correct effective action.  
  Another  equivalent way to state it is that one must maintain  the correct $2D$  Nambu-Goldstone direction as the scalar and gauge fields are rotated in the color-flavor \cite{GJK}. 

 As we noted above equations of motion (\ref{eqmdual1}) and (\ref{eqmdual2}) continue to hold as long as the derivatives $\de / \de x_{\alpha}$  
do not act on $\Pi_{B}$.  Here and below we again use the  indices $\alpha, \beta$  to indicate the coordinates on the worldsheet, whereas the indices $i,j$ are reserved for those in the plane perpendicular to the vortex length direction (see Fig.~\ref{Worldsh}).   Eq.~(\ref{eqmdual1}) for $(\mu, \nu)= (\alpha, \beta)$ contains the equations of motion for constant $B$  (plus corrections of at least second order in the derivative of $\Pi$).   Equations with $(\mu, \nu)= (i,j)$   are  of higher orders in    $\de_{\alpha},   \de_{\beta}$.  
Potential first-order corrections are in the   $(\mu,  \nu)= (\beta,  i)$ equation: 
\be    \de_{\alpha}  H^{\alpha \beta   i    \,  (B)}   +   \de_{j}  H^{j \beta   i    \,  (B)}      =  - m^{2} \,   G^{\beta  i  \, (B)}
\ee
 ($\Sigma^{\mu \nu}$ term is present only for $\mu, \nu = \alpha, \beta$ (see Eq.~(\ref{worldsheet})), or wrtten extensively  
\be \de^{\alpha}  [ \de_{\beta} G_{i  \alpha}^{(B)}  +  \de_{i} G_{\alpha \beta}^{(B)}  +  \de_{\alpha} G_{\beta i}^{(B)}  ]  + 
\de^{j}  [ \de_{\beta} G_{i  j}^{(B)}  +  \de_{i} G_{j \beta}^{(B)}  +  \de_{j} G_{\beta i}^{(B)}  ]    =  - m^{2} \,   G_{\beta i}^{(B)}\;.  \label{means}
\ee
The gauge field for static $B$ (``unperturbed'' solution with respect to $\de_{\alpha} \Pi_{B}$)     contains only 
\be   G_{\alpha \beta}^{(B)} = \tfrac{1}{2} \epsilon_{\alpha \beta ij} F^{(B)\, ij} =   \epsilon_{\alpha \beta}  \, {\cal B}^{(vor)}\Pi_{B},\qquad  \epsilon_{\alpha \beta}  =\begin{cases}
   1   &    \alpha \beta = 0 3 \\
     -1  & \alpha \beta = 3 0
\end{cases}    \label{remember}
\ee 
in the vortex region, far from the monopoles,  where   ${\cal B}_{i}= \delta_{i 3} {\cal B}^{vor}$  stands for the vortex magnetic field,   see Eq.~(\ref{using}).       Eq.~(\ref{means}) shows that new 
gauge components  $G^{(B)\, \beta  i }$  are generated, satisfying
\be   \de^{j}  [ \de_{i} G_{j \beta}^{(B)}  +  \de_{j} G_{\beta i}^{(B)}  ]   + m^{2}     G_{\beta  i}^{(B)} = -  \de^{\alpha}  \de_{i} G_{\alpha \beta}^{(B)}\; 
\label{this}\ee 
where we have dropped terms higher order in $\de \Pi_{B}$.

Eq.~(\ref{this}) can be solved for $G^{(B)\, \beta  i }$  by setting  
\be       G_{\beta i}^{(B)}  = \epsilon_{\beta i \alpha k}   R^{k} \, \de^{\alpha}  \Pi_{B}\;,    \label{newg}
\ee
and substituting it into Eq.~(\ref{this}) and recalling  Eq.~(\ref{remember}).  One finds after a simple calculation that 
\be    R^{i} =   \epsilon^{ij}   \,      
\frac{x_{j}}{ m^{2}  \rho} \de_{\rho}  {\cal B}^{(vor)}, \qquad   \rho =  \sqrt{x_{1}^{2} + x_{2}^{2}}, \qquad   \epsilon^{12}= - \epsilon^{21}=1\;.
\label{OK}\ee
where $ {\cal B}^{(vor)}$  is the usual  vortex magnetic field,    Eq.~(\ref{using}), Eq.~(\ref{using2}).      $G^{(B)\, \beta  i }$  can also be found more directly as follows.  
 We first convert equations of motion (\ref{eqmdual1}) and (\ref{eqmdual2}) into the form, 
 \be   \partial_\lambda \p^\lambda G^{\mu\nu\, (B)}  + m^2 G^{\mu\nu \,(B)} = \frac{m}{\sqrt{2}e}\Sigma^{\mu\nu\, (B)} - \frac{1}{\sqrt{2}em}(\p^\nu \p_\lambda \Sigma^{\lambda\mu \, (B)} - \p^\mu\p_\lambda \Sigma^{\lambda\nu \, (B)}). 
 \ee 
 For a static monopole and far from the monopole, we have
\begin{eqnarray}
 (\partial_i \p^i+ m^2) G^{\alpha\beta \, (B)} &=& \frac{m}{\sqrt{2}e}\Sigma^{\alpha\beta \, (B)}\;; \\
(\partial_j \p^j+ m^2) G^{\alpha i \, (B)}  &=& \frac{1}{\sqrt{2}em}\p^i \Sigma^{\alpha\beta}\p_\beta\Pi_{B}\;.
 \end{eqnarray}
 The solution for the first equation is 
 \begin{eqnarray}
G^{\alpha\beta\, (B)} = \frac{m}{\sqrt{2}e}K_0(m\rho) \Pi_{B}
\end{eqnarray} where $K_0$ is a modified Bessel function of the second kind. Solution to the second equation is then
 \begin{eqnarray}
G^{\alpha i\, (B)} = \frac{1}{m^2}\p^i G^{\alpha\beta}\p_\beta\Pi\;.  
\end{eqnarray}
in agreement with Eqs.~(\ref{newg}), (\ref{OK}).

Eq.(\ref{newg})  implies that the excitation energy of the vortex part is given by   
\be   \Delta E =    f\, \tr   \, ( \de_{\alpha}\Pi_{B} \de_{\alpha}\Pi_{B} ), \qquad 
  f  =  \frac{2\pi}{m^{2}}   \int_{0}^{\infty} d\rho  \,\rho \,   (\de_{\rho} {\cal B}^{(vor)})^{2}\;.   
\ee
The integral defining $f$ is logarithmically divergent at $\rho=0$, which must be regularized at the vortex width, $\rho \sim 1/ g v_{2}$.
By  a simple manipulation  (see e.g.  \cite{GJK})   this can be seen to be a $2D$  $CP^{N-1}$ action,  
\be   S^{2D}=   2f\, \int_{\Sigma}    d^{2} x\,   X^{-1}  \de^{\alpha} B^{\dagger}  Y^{-1} \de_{\alpha}  B\;, 
\qquad
(X\equiv 1  + B^\dag B \ , \quad
Y\equiv\mathbf{1}_{N-1}  + B B^\dag )\;,    \label{Bmodesagain}
\ee
defined on the worldstrip, $\Sigma$.
Note that the original $4D$ integration factorized into $2+2$,  because the $CP^{N-1}$ coordinate $B$ does not vary significantly over the range of the vortex width, $\sim 1/ gv_{2}$.

The monopole contribution to the effective action can be  found as follows.
Near a pointlike monopole,  $J^{0}=  \tfrac{4\pi }{g} \delta^{3}({\bf r}),$   and the magnetic field  $G^{0i}$   is the solution of 
Eq.~(\ref{eqmdual2}):  
\be          {\cal B}_{i}^{monopole}({\bf r})  =  \frac{1}{g} \, \de_{i}  \frac{e^{-  g v_{2} r}} {r}\;.
\ee   
Note that at distances larger than the {\it vortex }  width, $1/ g v_{2}$, this component is screened and dies out; the magnetic field $G_{0 i}$  is instead dominated  by the constant vortex  configuration.    
On the other hand, near the monopole this is a standard Coulomb field.   As  $J^{0}$ fluctuates in time in color-flavor,  
\be    J^0 \Pi \to    J^0 \Pi +  {\rm const.} \, J^0  \de_0 \Pi,   
\ee
  $G^{0i}$, $i=1,2,3$,   acquire a component in the $\de_0 \Pi$  direction, around the monopole, in order to maintain Eq.~(\ref{eqmdual2}) satisfied.  
It gives a singular contribution:
\be   \gamma \sim   \int d^{3}x  \,\sum_{i=1}^{3} \,  G_{0i} G^{0i} = \frac{1}{2 m^{2}}   \,  \int d^{3}x  \, \sum_{i}   ({\cal B}_{i}^{monopole}({\bf r}) )^{2} =  \frac{2\pi}{ m^{2} g^{2}}  
\int \frac{dr}{r^{2}},  \label{3Dint}  \ee
in the coefficient of  the fluctuation amplitude, $\de_0 \Pi  \, \de_0 \Pi$.
The singularity is regularized  at  the distances $\sim1/ gv_{1}$ where  the monopole turns smoothly into the regular  't Hooft-Polyakov configuration.
Therefore the integral in Eq.~(\ref{3Dint}) is dominated by the  radial region between  $1/ gv_{1}$ and   $ 1/ gv_{2}$,  over which the moduli parameter  $B({\bf r}, t)$ is regarded as constant. The $4D$ integration  here factorizes into  $4=3+1$.    The monopole contribution to the effective action is therefore 
\be    S^{1D}  =  \gamma \,  \int_{i= M, {\bar M}} dt  \,  X^{-1}   \de_{0} B^{\dagger} ({\bf r}_{i}, t)  Y^{-1} \de_{0} B ({\bf r}_{i}, t)\;,  \qquad \gamma  \sim   \frac{2\pi v_{1}}{ g^{3}v_{2}^{2}}\sim \frac{M}{m^{2}}\;, 
\ee
where $M= v_{1}/g$ is the monopole mass and $m=gv_{2}$ is the $W$ boson masses of the lower mass scale symmetry breaking.  

 The total effective action is a  $2D$  $CP^{N-1}$ theory with boundaries,     
 \be   S=  S^{2D} +  S_{M}^{1D} +  S_{\bar M}^{1D} \;.
 \ee
 There is a nontrivial constraint on the variable $B(x_{\mu})$:   
  on the boundary where the worldsheet meets the monopole worldline, the $CP^{N-1}$ variable matches:  
 \be     B(x(\sigma, \tau) ) |_{\sigma=0, \pi }  =   B_{M, {\bar M}}(t(\tau)) \;,  \label{boundary1} \ee
 or
\be      B(x_{3}, x_{0})|_{x_{3}= x_{M\, 3}}=  B_{M}(x_{0})\;;   \qquad   B(x_{3}, x_{0})|_{x_{3}= x_{M\, 3}}=  B_{M}(x_{0})\;.   \label{boundary2}
 \ee
This follows from  Eq.~(\ref{rotform}),   i.e., from the fact that the orientational zeromode of the monopole-vortex complex arises from the simultaneous $SU(N)_{C+F}$ rotations of the monopole and the light fields.

By introducing the complex unit $N$-component vector  $n^{c}$    ($c=1,2, \ldots, N$):
\be       n^{c}   =     
\begin{pmatrix}
X^{-\frac{1}{2}} \\
B X^{-\frac{1}{2}}
\end{pmatrix} =  
	\frac{1}{\sqrt{1 + B^{\dagger} B}} 
	\begin{pmatrix}
1 \\
B 
\end{pmatrix} \;,
 \label{identi} 
\ee
the vortex effective action above can be put into the familiar $SU(N)$ form of the $CP^{N-1}$ sigma model,  
\be        S^{2D}  =  2f\,  \int_{\Sigma} d^{2}x \,        {\cal D}_{\alpha}  n^{c\, \dagger} {\cal D}_{\alpha}  n^{c}, \qquad  
{\cal D}_{\alpha}  n^{c} \equiv \{ \de_{\alpha} -   (n^{\dagger}  \de_{\alpha} n) \} n^{c}, \qquad n^{\dagger} n =1
\ee
and similarly for the monopole action:
\be        S^{1D}  =   \gamma  \, \int_{K=M, {\bar M}}  d t  \,      {\cal D}_{0}  n_{K}^{c\, \dagger} {\cal D}_{0}  n_{K}^{c}, \label{heavy}
\ee
together with the boundary condition 
\be        n^{c}(x) |_{x= x_{M},  x_{\bar M}} =   n_{K}^{c} |_{K=M, {\bar M}}\;.   \qquad  \label{free}  
\ee
The boundary condition  Eq.~(\ref{boundary1}) or Eq.~(\ref{free}) can be thought of as something between Dirichlet  (in the infinite monopole mass limit, $\gamma  \to \infty$)  and   Neumann 
(in the light monopole limit), from the point of view of the $2D$   $CP^{N-1}$ model defined on the worldstrip of finite width \footnote{We thank Stefano Bolognesi for discussion on this point.}.

\section{Discussion}

Summarizing,  we have studied the vortex-monopole complex soliton configurations, in a theory with a hierarchical gauge symmetry breaking, so that the vortex ends at the monopole or antimonopole arising from the higher-mass-scale symmetry breaking.  The model studied has an exact color-flavor diagonal $SU(N)_{C+F}$  symmetry unbroken in the $4D$ bulk. The individual vortex-monopole soliton breaks it,  acquiring orientational
$CP^{N-1}$ zeromodes.  Their fluctuations are described by an effective $CP^{N-1}$ action defined on the worldstrip, the boundaries being 
the monopole and  antimonopole worldlines; in other words, the effective action is a $2D$ $CP^{N-1}$ model with boundaries, with the boundary condition, Eq.~(\ref{free}), plus the monopole 
$1D$ $CP^{N-1}$ action.
The boundary variable $n^{c}$ is a freely varying function of the worldline position, and acts as the source or sink of the excitation in the worldsheet.  

This illustrates the phenomenon mentioned in the Introduction.  Color fluctuation  of an endpoint monopole,  which in the theory without fundamental scalars   suffers from the non-normalizability of the associated gauge zeromodes \cite{DFHK} 
and would remain stuck (the famous failure  of the na\"{i}ve  nonAbelian monopole concept), escapes from the imprizonment as the color gets  mixed with flavor in a color-flavor locked vacuum, and propagates freely
  on the vortex worldsheet.   In the dual description the monopoles appear as pointlike objects, transforming under the fundamental representation of this new $SU(N)$ symmetry - the isometry group of the $CP^{N-1}$ action.  It is a local $SU(N)$ symmetry, albeit in a confinement phase: these fluctuations 
do not propagate in the bulk outside the worldstrip.    The $M-V-{\bar M}$ system as a whole is a singlet of the new $SU(N)$.     This is appropriate because the original color $SU(N)$ is in the Higgs phase. Its dual must be in a confinement phase. 

We see now  how a nonAbelian dual $SU(N)$ system emerges, not plagued by any of the known problems.  The so-called topological obstruction is cured here, as the bare Dirac string singularity of the monopole, which lies along the vortex core,  is eaten by the vortex, so to speak.  The scalar field vanishes along the vortex core, and precisely cancels the singularity in the action.   This is most clearly seen in the explicit microscopic description of the monopole-vortex complex such as in \cite{CDGKM}.  

Let us end with some more remarks. 

Magnetic monopoles have also been studied in the context of a $U(N)$ theory in a color-flavor locked vacuum (i.e., with $N_{f}=N$ number of flavors), {\it  but without the underlying $SU(N+1)$ gauge theory} \cite{Shifman:2004dr,Hanany:2004ea,Gorsky:2004ad,NittaVinci,Massdeformed}.  By choosing unequal masses for the scalar fields, 
$m_{i}\ne m_{j}$ the flavor (and hence color-flavor) symmetry is explicitly broken, and degenerate $N$ Abelian vortices appear, instead of continuous set of nonAbelian vortices, parametrized by $CP^{N-1}$ moduli. Monopoles appear as kinks connecting different vortices, having masses of order of $O(|m_{i}-m_{j}|/g)$.
These are Abelian monopoles.  In order to find candidate nonAbelian monopoles in such a context, one must choose judiciously the scalar potential (partially degenerate) \cite{NittaVinci,Massdeformed,Monin:2013mxa}  so that one finds in the same system degenerate vortices of Abelian and nonAbelian types.   

In the limit of equal masses  $m_{i}= m_{j}$, the semiclassical analysis above is no longer reliable.  But since in these systems the vortex is infinitely long (stable), 
one can make use of the facts known about the infrared dynamics of $2D$ $CP^{N-1}$ theory.  It is in fact known that the quantum fluctuations of the $CP^{N-1}$ modes
become strongly coupled at long distances (a $2D$ analogue of confinement) \cite{Witten,Shifman}; it means that the vortex dynamically Abelianizes \cite{ABEKY,Shifman:2004dr,Hanany:2004ea,Gorsky:2004ad}.  The masses of the kink monopoles are now replaced by $O(\Lambda)$, where $\Lambda$ is the dynamical scale of the $2D$ $CP^{N-1}$ theory.    In particular, in the case of an ${\cal N}=2$ supersymmetric  model, the effective $2D$ theory on the vortex world sheet is a  $(2,2)$ supersymmetric  $CP^{N-1}$ model.   Quantum effects lead to $N$ degenerate vacua ($N$ Abelian vortices); monopoles appear as kinks connecting adjacent vortices.  A close connection of these objects and the $4D$  (Abelian) monopoles appearing in the infrared in  $4D$,  ${\cal N}=2$ supersymmetric gauge theories has been noted \cite{Shifman:2004dr,Hanany:2004ea,Gorsky:2004ad}, which seems to realize the elegant $2D$-$4D$ duality proposal made earlier by N. Dorey \cite{Dorey}.  These monopoles are confined by two Abelian vortices \cite{Gorsky:2004ad}, in contrast to the monopoles considered in the present work.

Thus even though our system below the mass scale $v_1$ has some similarities as those considered in \cite{Shifman:2004dr,Hanany:2004ea,Gorsky:2004ad}, 
they are clearly physically distinct.  Our vortex has the endpoint monopoles, whose properties have been our main interest.  In fact, the effective world sheet $CP^{N-1}$ action found here is defined on a finite worldstrip, with endpoint monopoles having their own $CP^{N-1}$ dynamics.  It is an open problem what the infrared dynamics of the $CP^{N-1}$ system defined on such a finite-width worldstrip with the boundary condition Eq.(\ref{free}) is,  and  how the low-energy phase depends on the width of the worldstrip (the vortex length).\footnote{A $CP^{N-1}$ model with a Dirichlet boundary condition and at large $N$,   was studied recently  \cite{Milekhin}.   It shows a phase transition from Higgs to confinement phase  at a critical vortex length. }
  
  The metastability of our vortex-monopole system also means that, when one tries to stretch the vortex it will be broken by  spontaneous creation of a monopole-antimonopole pair. 
In this sense the vortex length  itself  is also a dynamical variable,  dependent on the ratio   $v_{2}/v_{1}$.  

Another issue to be kept in mind is the possible relevance of hierarchical symmetry breaking but with reduced gauge and flavor symmetry at the first stage $v_{1}$, such as $SU(N+1) \to  SU(r) \times SU(N-r)\times U(1)$.  In that case the soliton vortex-monopole system carries orientational moduli of a product form,  $CP^{r-1}\times CP^{N-r-1}$.  It is possible that in such a system dynamical Abelianization occurs only partially \cite{DKO}, reminiscent of the quantum $r$ vacua of the ${\cal N}=2$ SQCD. 

More generally, the monopole and antimonopole positions must also be treated as soliton collective coordinates and their motion should be taken into account as an additional piece  to the action.   In Eq.~(\ref{heavy}) we assumed that the monopoles are very heavy ($v_{1}\gg v_{2}$) and do not move appreciably; taking their motion into account introduces a space variable dependence of the monopole  variable, $n_{K}(x_{0}) \to n_{K}(x_{0}, x_{3})$,  $K= M, {\bar M}$, and the $CP^{N-1}$ dynamics and the space motion of the monopole positions will get mixed (see for example \cite{CiprianiF})).  

In the large-distance approximation we have adopted, the $U(1)$ moduli of the classical 't Hooft-Polyako monopole solutions - which rotates the exponentially damped part of the configuration \cite{EW} -  is not seen. The (internal)  monopole moduli ($CP^{N-1}$ rather than $CP^{N-1}\times S$)  coincides with that of the vortex attached to it.    Of course, the electric charge of the monopole due to  Witten's effect is correctly taken into account in our large-distance approximation, see Appendix,~\ref{metric}.

A final remark concerns the flavor quantum numbers of the monopole, arising from e.g., the 
Jackiw-Rebbi effect \cite{JR}.  In the case of a supersymmetric extension of the model considered here (softly-broken ${\cal N}=2$ SQCD), due to the fermion zeromodes associated with each scalar $q$ in the fundamental representaiton of $SU(N+1)$,   the monopole acquires flavor global charge.  Due to the normalizability of the associated fermion $3D$ zeromodes,  
this effect is localized near the monopole center (of distances $\sim 1/v_{1}$).  Its fluctuation does not propagate, and is clearly distinct from the role played by the flavor symmetry at large distances $\sim1/v_{2} \gg  1/v_{1}$  in generating the dual local  $SU(N)$ system via the color-flavor locking.    The flavor quantum number of the monopole is, in turn,  fundamental in the renormalization-group behavior in the dual theory. 

Global flavor symmetry thus plays several key roles,  intertwined with soliton and gauge dynamics,  
 in generating local dual nonAbelian symmetry.

\section*{Ackonowledgment}  We thank Stefano Bolognesi, Simone Giacomelli, Sven Bjarke Gudnason, Muneto Nitta, Keisuke Ohashi and Norisuke Sakai for useful discussions and 
Keio University for hospitality.

  \appendix

\section{Minimizing the potential \label{minimum}}  

The first term of Eq.~(\ref{potential}), after going to the matrix representation of the adjoint field ($\phi \equiv \phi^{A} T^{A}$)  and  by using the Fierz relation (for $SU(N+1)$)
\be (T^{A})^{a}_{b} (T^{A})^{c}_{d}  =     -\frac{1}{2(N+1)}   \delta^{a}_{b}  \delta^{c}_{d} +\frac{1}{2}   \delta^{a}_{d}  \delta^{c}_{b},  \ee 
reads
\be   V_{\eta}=    \Tr\, \left| \mu \phi  -   \frac{1}{2 (N+1)} (\sum_{I}{q}^{\dagger \, a}_{I}  q^{I}_{a} ) {\mathbbm 1} +  \frac{1}{2}  q \, q^{\dagger} \right|^{2} \;. 
\ee
By dropping the massive $\chi$ fields, and by using the decompositions, Eq.~(\ref{Mfiel}) and Eq.~(\ref{decompos}),  this becomes 
\be   {q}^{\dagger \, a}_{I}  q^{I}_{a}  =  \eta^{\dagger}_{I}  \eta_{I}, \qquad   q^{a}_{ I}  {q}^{\dagger}_{b\, I}     =    e^{a}_{i}\, {\bar e}^{j}_{b}  \, \eta^{i}_{I}  \eta^{\dagger}_{j  \, I}
\equiv  e {\bar e} \eta \eta^{*}  \;;
\ee
\be   V_{\eta}=\Tr  \left|    {\mu m_{0} (N+1)}  (z {\bar z}-  \frac{1}{N+1}  {\mathbbm 1})  -  \frac{1}{2 (N+1)}   (\eta^{\dagger}_{I} \eta_{I})   {\mathbbm 1}   + \frac{1}{2} e {\bar e} \eta \eta^{*} \right|^{2}. 
\ee
By using the completeness 
\be    z{\bar z} + e {\bar e}= {\mathbbm 1},  
\ee
\bea  V_{\eta} &=& \Tr \left|\left(  {\mu m_{0} N }   -   \frac{1}{2 (N+1)}   (\eta^{\dagger}_{I} \eta_{I})   \right)  z {\bar z}  -  \left(  {\mu m_{0} }   +   \frac{(\eta^{\dagger}_{I} \eta_{I})  }{2 (N+1)}      \right) e {\bar e}    + \frac{1}{2}  e {\bar e} \eta \eta^{*} \right|^{2}. 
\nonumber \\
&=&     \left({\mu m_{0} N }   -   \frac{1}{2 (N+1)}   (\eta^{\dagger}_{I} \eta_{I})  \right)^{2} +  
\Tr \left|   \left(   {\mu m_{0} }   +   \frac{(\eta^{\dagger}_{I} \eta_{I})  }{2 (N+1)}      \right) e {\bar e}      - \frac{1}{2}  e {\bar e} \eta \eta^{*}    \right|^{2}  \label{secondt}
\eea
The minimum of the first term gives, 
writing 
\be   \eta^{\dagger}_{I}  \eta_{I}  \equiv    N\, d^{2}, \ee
\be  N^{2} \left( {\mu m_{0} }  -     \frac{  d^{2} }{2(N+1)}  \right) =0, \qquad .^{.}. \quad
    d^{2}=   2 (N+1) \mu m_{0}\,.\label{thesame}
\ee
   As for the second term,
one has, by using 
\be    \Tr   (e_{i} {\bar e}^{i}) (e_{j} {\bar e}^{j})  = N, \qquad  \Tr   (e_{i} {\bar e}^{i})   ( e_{j} {\bar e}^{k} \eta_{I}^{j} \eta^{I\,  *}_{k} ) =   \eta^{\dagger}_{I}  \eta_{I} =  N\, d^{2}, \qquad 
 \ee
 \be \Tr    ( e_{j} {\bar e}^{k} \eta^{j}_{I} \eta^{*}_{k\, I} )  ( e_{\ell} {\bar e}^{m} \eta^{\ell}_{J} \eta^{*}_{m \, J } ) =  \delta^{k}_{\ell}\delta^{m}_{j} \eta^{j}_{I} \eta^{*}_{k\, I}  \eta^{\ell}_{J} \eta^{*}_{m \, J }= (\eta^{\dagger}_{I} \eta_{J})   (\eta^{\dagger}_{J} \eta_{I}), 
\ee
and the second term of Eq.~(\ref{secondt}) becomes 
\be  {N}\,   \left[  \left( {\mu m_{0} }   +   \frac{ N d^{2} }{ 2(N+1)} \right)^{2}  -    \left( {\mu m_{0} }   +   \frac{ N d^{2} }{ 2(N+1)} \right)  d^{2}  +  \frac{d^{4}}{4}    \right]  =  
 \frac{N}{2}\,     \left( {\mu m_{0} }   -  \frac{  d^{2} }{ 2 (N+1)} \right)^{2} 
\ee
which gives the same condition  as Eq.~(\ref{thesame}).    Eq.~(\ref{secondt})  leads also to the conditions
\be     \sum_{I\ne J}  | \eta^{\dagger}_{I} \eta_{J}|^{2} =0, \quad   .^{.}.   \quad   \eta^{\dagger}_{I} \eta_{J}=0, \quad   I\ne J, \qquad    
\eta^{\dagger}_{1} \eta_{1}=\eta^{\dagger}_{2} \eta_{2}=\ldots = \eta^{\dagger}_{N} \eta_{N}.
\ee
These imply that  
\be     \bra  \eta^{i}_{I}  \ket= \delta^{i}_{I}   \, \sqrt{2(N+1) \mu m_{0}} \equiv    v_{2}\, \delta^{i}_{I}\;.
\ee

\section{Monopole and vortex flux matching}  

The fact that the monopole magnetic flux is precisely whisked away by the vortex attached to it in the context of the hierarchical symmetry breaking, has been carefully studied
by Auzzi et. al. \cite{ABEK}, and we recall simply the results in our setting. Although the monopole is coupled to the light scalars through the field $P_{\mu}$  (see Eq.~(\ref{action})), the latter is  a part of the 
the $su(2) \subset su(N+1)$ field 
\be     B_i^a \frac{\tau^a}{2}  =  B_i^3 \frac{\tau^3}{2} = \frac{1}{2g}   \nabla_i \frac{1}{r} \left(\begin{array}{cc}1 & 0 \\0 & -1\end{array}\right)\;;
\ee
the magnetic flux through a tiny sphere around the monopole is given by 
\be     \int  d{\bf  S}\cdot {\bf B} =     - \frac{2\pi}{g}  \tau^3\;,  \qquad   \int  d{\bf  S}\cdot {\bf B^3} =   - \frac{4\pi}{g}\;.      \label{exactly}
\ee
The vortex  gauge field $A_{\mu}$ is  such that the winding  of the scalar field is cancelled at large distances from the vortex center, in the 
matter kinetic term, 
\be     |  (  \de_\mu -  i g  A_{\mu}   ) q  |^2\,.    
\ee 
In the gauge   where the light field $\eta$ has the form,  (Eq.~\ref{particular1}),  the $(1 1)$ component of the $SU(N)$ gauge field has the asymptotic behavior
\be     A_{\phi}   \sim  -  \frac{1}{ g  \rho}\;.  
\ee
This must a part of the traceless $SU(N+1)$ gauge field
\be   A_{\phi} \sim  \tau^3      \frac{1}{ g  \rho} \;,   
\ee
belonging to $U(1) \subset SU(2) \subset  SU(N+1)$.
The vortex flux through a plane perpendicular to the vortex axis is then
\be      \int d^2x  \,   \nabla \times  A_{\phi}  =   \oint    dx_i     A^i =    \int \rho \,d\phi \,     A_{\phi}   =    \frac{2\pi}{ g}  \tau^3\;,
\ee
which matches exactly the monopole flux, Eq.~(\ref{exactly}).

\section{Solution of the dual equations of motion \label{metric}}

In the presence of the static (heavy) monopoles at
\be   {\bf r}_{1}= {\bf r}(\tau, 0) = (0,0, - z_{1}), \qquad   {\bf r}_{2}= {\bf r}(\tau, \pi) =(0,0,0),    \label{monopolesBisbis}\ee  
the worldstrip is at
\beq   \Sigma^{30}=-\Sigma^{03} =  2 \pi  \delta(x)\delta(y) \theta(-z) \theta(z+ z_{1})\;, \qquad    \Sigma^{\mu \nu}=0 \quad (\mu \nu) \ne (30), (03)\;; \eeq 
which clearly satisfies the monopole confinement condition (\ref{builtin}).  The equation of motion for $G_{\mu \nu}$ has been solved in \cite{KMO}.

 In order to interpret the result in terms of the original electric and magnetic fields, we note that the duality transformation  implies ($\alpha  \equiv    {\theta g^{2} }/{ 8 \pi^{2}}$):
\bea   F_{\mu \nu} &=&  -\frac{m}{1 + \alpha^{2}}    ({\tilde G}_{\mu \nu} - \alpha G_{\mu \nu})   =  - \frac{1}{g} {\tilde \Sigma}_{\mu \nu}  -  \frac{1}{m}  (\de_{\mu}  L_{\nu} - \de_{\nu}  L_{\mu}) \;  \nonumber \\
&=&     - \frac{1}{g} {\tilde \Sigma}_{\mu \nu}  -      \frac{1}{\Box +m^{2}}    \left[  \de_{\mu} (\alpha  j_{\nu} + {\tilde j}_{\nu} )  -   (\mu \leftrightarrow  \nu)      \right]\;.
 \label{elmagfield} \eea

For instance, let us consider a massive static monopole sitting at ${\bf r}=0$ with a vortex attached to it and extending into the $-{\hat z}$ direction:  
\beq   \Sigma^{30}=-\Sigma^{03} =4 \pi \, \delta(x)\delta(y) \theta(-z)\;, \qquad    \Sigma^{\mu \nu}=0 \quad (\mu \nu) \ne (30), (03)\;; \eeq 
\beq   j^{0}  =   \frac{ 4 \pi }{g}\, \delta^{3}({\bf r}),  \quad  j^{i}=0\;; \quad i=1,2,3\;;  \qquad   {\tilde j}^{\nu} =  -\frac{1}{g}      \epsilon^{\lambda  \nu  0  3}  \, \de_{\lambda} \Sigma_{03}\;.
\eeq
 From Eq.~(\ref{elmagfield}) one finds that (we recall $\alpha= \theta g^{2}/  8 \pi^{2}$)
 \beq     {\cal E}_{i} =  F_{0i}=   \alpha\, {\cal B}_{i}^{(mon)}, \qquad {\cal B}_{i} = \frac{1}{2} \epsilon_{ijk} F_{jk}=     {\cal B}_{i}^{(mon)} +   {\cal B}^{(vor)}   \delta_{i}^{3}\;,  \label{remark}
 \eeq
 where 
 \beq      {\cal B}_{i}^{(mon)} =  \frac{1}{g}  \de_{i}  G({\bf r}), \qquad  {\cal B}^{(vor)}= \frac{1}{g} \,   m^{2} \,  \int_{-\infty}^{0} \, dz^{\prime} \,  G(x, y, z-z^{\prime})\;,\label{using}
 \eeq
 and $G({\bf r})$ is the Green function, having the Yukawa form
 \beq     G({\bf r}) = \frac{4\pi}{- \Delta + m^{2}}\, \delta^{3}({\bf r}) =     \frac{e^{-m r}}{r}\;.  \label{using2}
 \eeq
   Note the clear-cut separation of the monopole and vortex contributions to magnetic (and electric) fields,  Eq.~(\ref{remark}).  
   In the vortex region,  the only nonvanishing component is the magnetic field in the $x_{3}$ (vortex length) direction, $F_{12}\sim  G_{03}\sim  {\cal B}^{(vor)}$.

\end{document}